\documentclass{IEEEtran}
\usepackage{graphicx}
\usepackage{subfigure}
\usepackage{amsmath}
\usepackage{balance}
\usepackage{float}
\usepackage[misc,geometry]{ifsym}
\usepackage{color}
\usepackage{url}
\usepackage{verbatim}
\usepackage{xspace}
\usepackage{array}
\newcolumntype{P}[1]{>{\centering\arraybackslash}p{#1}}
\usepackage[utf8]{inputenc}
\usepackage{multirow}
 
\begin{document}

\title{Blockchain Testing: Challenges, Techniques, and Research Directions}

\author{\IEEEauthorblockN{Chhagan Lal\IEEEauthorrefmark{1}, and
Dusica Marijan\IEEEauthorrefmark{3}}\\
\IEEEauthorblockA{Simula Research Laboratory, Oslo, Norway\IEEEauthorrefmark{1}\IEEEauthorrefmark{2}\\
Email: \IEEEauthorrefmark{1}chhagan.iiita@gmail.com,
\IEEEauthorrefmark{2}dusica@simula.no}}

\maketitle

\begin{abstract}

Specific testing solutions targeting blockchain-based software are gaining huge attention as blockchain technologies are being increasingly incorporated into enterprise systems. As blockchain-based software enters production systems, it is paramount to follow proper engineering practices, to ensure the required level of testing, and assess the readiness of the developed system. The existing research aims at addressing the testing related issues and challenges of engineering blockchain-based software by providing suitable techniques and tools. However, like any emerging discipline, the best practices and tools for testing blockchain-based systems are not yet sufficiently developed. In this paper, we provide a comprehensive survey on testing of Blockchain-based Applications (BC-Apps). First, we provide a discussion on identified challenges that are associated with BC-App testing. Second, we use a layered approach to discuss the state-of-the-art testing efforts in the area of BC technologies. In particular, we present an overview of the existing testing tools and techniques that provide testing solutions either for different components at various layers of the BC-App stack or across the whole stack. Third, we provide a set of future research directions based on the identified BC testing challenges and gaps in the literature review of existing testing solutions for BC-Apps. Moreover, we reflect on the specificity of BC-based software development procedure, which makes some of the existing tools or techniques inadequate, and call for the definition of standardised testing procedures and techniques for BC-Apps. The aim of our study is to highlight the importance of BC-based software testing and to pave the way for a disciplined, testable, and verifiable BC software development.
\end{abstract}


\IEEEpeerreviewmaketitle

\section{Introduction}\label{sec:introduction}
In recent years, it has been seen that the usage of blockchain (BC) technology is not limited just to the financial sector as it was the case when it first gained popularity by playing a vital role in the Bitcoin system~\cite{bitcoin}. The researchers in academia and industry are currently increasingly investigating the usage applications of BC technology across many domains, ranging from manufacturing and healthcare to insurance and aeronautics~\cite{DIFRANCESCOMAESA202099}. It is mainly because of BC's inherent features like decentralization, immutability, improved security, transparency, and its ability to securely implement complex business logic and processes through smart contracts (SCs). On the one hand, these very features open new business opportunities and provide various advantages to businesses~\cite{8642861}~\cite{9179702}~\cite{9007406}, but also add new dimensions and standpoints that require specific attention from the testing community~\cite{8529728}. Therefore, with this increasing demand to leverage the BC technology for various purposes in different applications, it is equally important to perform adequate testing for potential bugs and vulnerabilities that could lead to multiple security threats or asset losses~\cite{praitheeshan2019}. In particular, along with the increasing deployment and integration abilities between BC and real-world applications, a testing strategy to effectively and efficiently test BC-Apps is paramount. Moreover, it is important to identify the components and interfaces in BC-Apps for which performing adequate testing is critical. 
\par BC-Apps significantly differ from other applications (i.e., applications not using BC technology). Therefore, they have specific requirements and acceptance criteria to be achieved during testing~\cite{8529728}~\cite{7965292}. For instance, in BC, the SCs have high significance because their execution cannot be reversed once implemented. Due to BC's immutable nature, once a buggy SC goes into a production system, it might need a complete revision of the code. Besides, the SC code also defines how efficiently the software performs with increasing workloads. Therefore, it becomes necessary to perform comprehensive performance testing to ensure that SCs work correctly. SC testing should be done as early as possible, i.e., when SCs are developed and not yet deployed. Moreover, most of the SCs are written in new programming languages (e.g., Solidity~\cite{3103305}, and Go~\cite{2851099}), for which either the testing tools are not available or not at mature stages. For BC-Apps, the correct functionality of SCs is paramount, because bugs in SCs may lead, and indeed have lead, to immense asset losses and disruptions~\cite{5445568}. Bugs even occurs in SCs written by experienced programmers, which highlights the fact that SC programming is challenging~\cite{8847638}. Therefore, to support the development of secure SCs, numerous tools and techniques have emerged in recent years~\cite{praitheeshan2019}~\cite{SINGH2020101654}. Moreover, some existing works also help the analysis of already deployed SC analysis tools~\cite{8732934}.

\par Apart from SCs, the decentralized and anonymous nature of the participating nodes that work together with distributed systems in a peer-to-peer network further adds to the BC testing complexity. For instance, the distributed nature enforces the need for the validation of synchronization between the nodes, and testing the performance and security of consensus algorithms~\cite{ferdous2020blockchain}. Consequently, BC technology introduces an entirely new perspective to software development, and validating and verifying a BC implementation's correctness poses certain new challenges compared with non-BC software. Therefore, the existing testing techniques or strategies may not be valid anymore, which calls for specialized testing tools and practices. Apart from the knowledge of rigorous testing protocols, developers building BC-based software need other competencies, including the understanding of various blockchain technologies and frameworks, and the understanding of different cryptographic techniques and consensus algorithms~\cite{BAMAKAN2020113385}. Moreover, the BC-App developers need specialized testing capabilities to perform testing for components (SCs, peer nodes, and consensus algorithms) that are new and specific to BC-based systems.  

\begin{table*}
\caption {Comparison with the related works}
\centering
\scalebox{1.2}{
\begin{tabular}{|P{2cm}|P{2cm}|P{2cm}|P{2cm}|P{2cm}|P{2cm}|}\hline
\textbf{} & \textbf{SC Testing} & \textbf{Performance Testing} & \textbf{Security Testing} & \textbf{API and Interface Testing} & \textbf{BC Testing Challenge Identification }\\ \hline

\cite{8327567}, 2018 & Yes & No & No & No & Partially \\ \hline

\cite{8732934}, 2019 &
Yes &
No &
No &
No &
No \\ \hline
\cite{8782988}, 2019  &
Yes &
No &
No &
No &
No \\ \hline
\cite{praitheeshan2019}, 2020  &
Yes &
No &
No &
No &
No \\ \hline
\cite{9129732}, 2020 &
No &
Yes &
No &
No &
No \\ \hline
\cite{9239372}, 2020 &
Partially &
No &
Yes &
No &
Partially \\ \hline
\cite{9271868}, 2020 &
Partially &
No &
Yes &
No &
Partially \\ \hline
This survey &
Yes &
Yes &
Yes &
Yes &
Yes 
\\ \hline

\end{tabular}}
\label{related}
\end{table*}

\par Since SCs are the most critical entities in a BC implementation, a large percentage of the state-of-the-art research work is performed in the area of testing SCs for potential bugs and security vulnerabilities. To test SCs, researchers have used different approaches that include formal verification methods (e.g., verification methods, model checking, and theorem proving)~\cite{3386022}~\cite{SINGH2020101654}, fuzzing methods (e.g., mutation, and hybrid)~\cite{3238177}~\cite{torres2020smart}, automated program repair~\cite{yu2020SCRepair}, symbolic execution and analysis~\cite{8952204}, and Control Flow Graph (CFG) construction~\cite{2976749297}. All these tools/techniques aim to automate error detection process while maximizing code coverage and the number of vulnerabilities detected. Other than SC testing, the other types of testing that attracted quite a few research works include performance testing, where the performance of the target BC-based system under test (SUT) is evaluated against different workloads and faultloads~\cite{8946222}~\cite{inbook1007}. Apart from SC and performance testing, other testing areas, such as peer/node testing, API testing, and consensus algorithm's testing, are not adequately explored by the research community. Due to the challenges inherit to BC testing, such as the lack of testing approaches, and the financial and data protection risks associated with errors in BC-Apps, it is of utmost importance that the researchers explore this topic: (i) to identify the testing-related challenges for BC, the key system components that need to be tested, and the ongoing efforts on performing testing, along with their limitations, and (ii) to design novel quality assurance (QA) strategies and testing techniques to address the identified issues. 

\par \textbf{Contribution.} There are several survey articles available in the state-of-the-art on testing-related efforts of BC technology. However, these articles cover only the partial aspects of BC testing, for example, SC testing~\cite{8327567}~\cite{8732934}~\cite{8782988}~\cite{praitheeshan2019} or BC performance testing~\cite{9129732} or security testing~\cite{9239372}~\cite{9271868}. To the best of our knowledge, there is no previous study providing a comprehensive survey of different testing techniques for BC components at different layers and across the whole BC stack. Moreover, existing surveys do not consider the testing of BC technology from the viewpoint of its integration with various real-world applications (i.e., standalone aspects of BC testing are studied instead of its coexistence with a target application). Finally, existing surveys do not investigate the new testing challenges that BC-App developers and testers face while developing and testing BC software. Understanding these challenges is an essential step towards progressing the current state-of-the-art on BC testing. Table~\ref{related} provides a comparison of our survey with the state-of-the-art by considering several parameters. The comparison clearly shows the need for our study in the domain of BC-App testing.   

The key contributions of our paper are as follows.

\begin{itemize}
    
\item We provide a comprehensive discussion on the importance of BC-Apps testing. We also discuss how it differs from testing the non-BC based implementations and what new challenges or issues arise when dealing with SCs and distributed applications (aka Dapps) in the scope of BC-Apps. 

\item We present a detailed survey of existing techniques and tools investigated and developed to perform different types of testing in BC-based systems. We also identify the limitations of these testing solutions, and we provide a comparison between different testing approaches by discussing their efficiency, performance, and practicability. To the best of our knowledge this is the first study that considers, in detail, the testing related aspects of the BC-Apps.

\item Using a layered design for BC-Apps, we discuss possible approaches to identify the key components at each layer, and the interfaces across different layers, that need to be tested along with the best techniques to test them. Finally, we provide several research directions to help prospective researchers working in this domain to gain technical insights, such as learning secure programming practices, create a mindset on test-centric development strategy, and be ready to collaborate in the community of diverse motivation. Moreover, they should also prepare themselves by acquiring knowledge in distributed networking and programming, cryptography, and mathematics.
\end{itemize}

The remaining of this paper is organized as follows. We provide the required background that includes a brief discussion about BC and SC technology, along with their key benefits, the architecture of BC-Apps depicting the major components and their interactions, and an overview of testing techniques and types used in software engineering disciplines in Section II. The key testing-related challenges for BC-Apps are presented in Section III. In Section IV, the state-of-the-art testing tools and techniques for performing different types of testing on various BC components is provided in detail. Section V includes a discussion based on the BC-Apps testing challenges and state-of-the-art techniques along with the directions of future research work in the area. Finally, we conclude our work in Section VI.      

\section{Background}
In this section, we present a brief discussion on the topics like BC, smart contracts, BC-Apps, and testing techniques that are relevant to fully understand the contributions that we make in the later sections.  

\subsection{Blockchain and Smart Contracts}
BC is a distributed ledger consisting of a series of chronologically ordered blocks appended in a link-list type of data-structure. To provide integrity and immutability of data in the ledger, it is required to prevent any updates in the committed blocks. To ensure this, each block contains the hash of the previous block, and the ledger is replicated across peers in the BC network. A block usually contains a set of timestamped transactions that are bundled together and stored in the form of a Merkle tree~\cite{704751}. BC adopts various cryptographic primitives like hashing algorithms, digital signatures, and PKI protocols to ensure adequate security. There are two key participants in the BC network, one that generates the transactions and the other that validate and store them in the ledger. 
\par A BC network runs on a peer-to-peer topology where each node is expected to store the same copy of the ledger. It consists of nodes or organizations that do not have a preexisting trust relationship among them. Therefore, to ensure that each peer node has the same copy of the ledger at any given time, the new valid block that will be appended in the ledger is selected by executing a consensus mechanism. In particular, a consensus mechanism, e.g., Practical Byzantine Fault Tolerance (PBFT), Proof-of-Work (PoW), and Proof-of-Stake (PoS), is a protocol that ensures synchronization among all network peers about the validity and ordering of transactions~\cite{8972381}. Therefore, these mechanisms are pivotal for BC's correct functioning and need to be tested properly before their use in real-world applications. The key components and functionalities of a BC-based system enable some unique features including immutability, decentralization, consensus, provenance, and finality. These features make BC an ideal solution, not only for the financial applications but also for non-financial ones~\cite{9184022}~\cite{8735815}.

\par A BC can be public (aka permissioned) or private (aka permissionless) type. A public BC (e.g., Bitcoin) allows anyone to become a participant, which means they can perform activities like taking part in the consensus mechanism, sending new transactions in the network, and maintaining a ledger state. While in a private BC (e.g., Hyperledger Fabric), the participation is constrained, and only the pre-verified parties known to each other are allowed to join the network. The choice of BC type depends on the target use case application for which it is being deployed. For instance, if a network can provide trust between the participating entities without requiring their verification by using PoW or PoS approaches, then the use of a public BC makes sense. On the other hand, an application such as medical data management in the healthcare domain which requires that all participating entities are vetted before accessing the medical records, due to its sensitive nature, makes it an ideal use case for a private BC~\cite{9091543}. In particular, when the access control and identity management procedures are vital for all the participants to execute transactions and others to see its origin, a private BC makes sense. Please note that there are also other benefits and drawbacks, apart from the trust among the participants, for each type of BC. They include scalability~\cite{8823874}, security and privacy~\cite{3316481}, and degree of decentralization, which should be taken into account when making a choice between private and public BC. Since the detailed description of BC technology is out-of-the-scope of this paper, we direct the interested readers to other comprehensive research works on this area, such as~\cite{8760539}, ~\cite{8369416}, and ~\cite{3366370}.

\par A BC can use Smart Contracts (SC), which are securely stored on the BC and are executed manually (via a transaction invoking a function in it) or automatically (when a precondition evaluates to true). Specifically, it allows for decentralized automation by facilitating the verification and enforcement of conditions written in the underlying contract~\cite{ethereum}. In this way, SCs serve as agreements or a set of policies that supervise a transaction. For instance, an SC can define a set of rules for an individual’s travel insurance, which trigger’s the contract execution when a travelling carrier (such as flight or train) is experiencing delay by more than a fixed amount of time. In particular, an SC consists of a set of instructions or operations written in special programming languages (e.g., Solidity and Go Lang), and it gets executed upon the fulfillment of predefined conditions. The key property that makes SCs of great use in many real-world applications is their ability to eliminate the requirement of a trusted third party in multiparty interactions~\cite{MACRINICI20182337}. Parties can participate by performing secure peer-to-peer transactions over BC without placing their trust in outside parties that are generally used to ensure that all parties fulfill the contractual obligations. Examples of application areas where SCs can provide several benefits are medical data sharing and management, escrow, supply-chain management, and e-voting. Moreover, with the advancements in the Internet of Things (IoT) domain, SCs could also play a vital role in machine-to-machine communications~\cite{8600854}. Currently, the largest BC platform for SCs deployment is Ethereum~\cite{ethereum}. It uses Solidity, a high-level scripting language that is specifically designed to write SCs~\cite{3103305}. Solidity is inspired by common programming languages like C++, JavaScript, and Python. Thus, it supports properties such as inheritance, libraries, and user-defined types, which helps the developers to depict complex business logic using SCs. 

\begin{figure}
\centering
  \includegraphics[scale = .105]{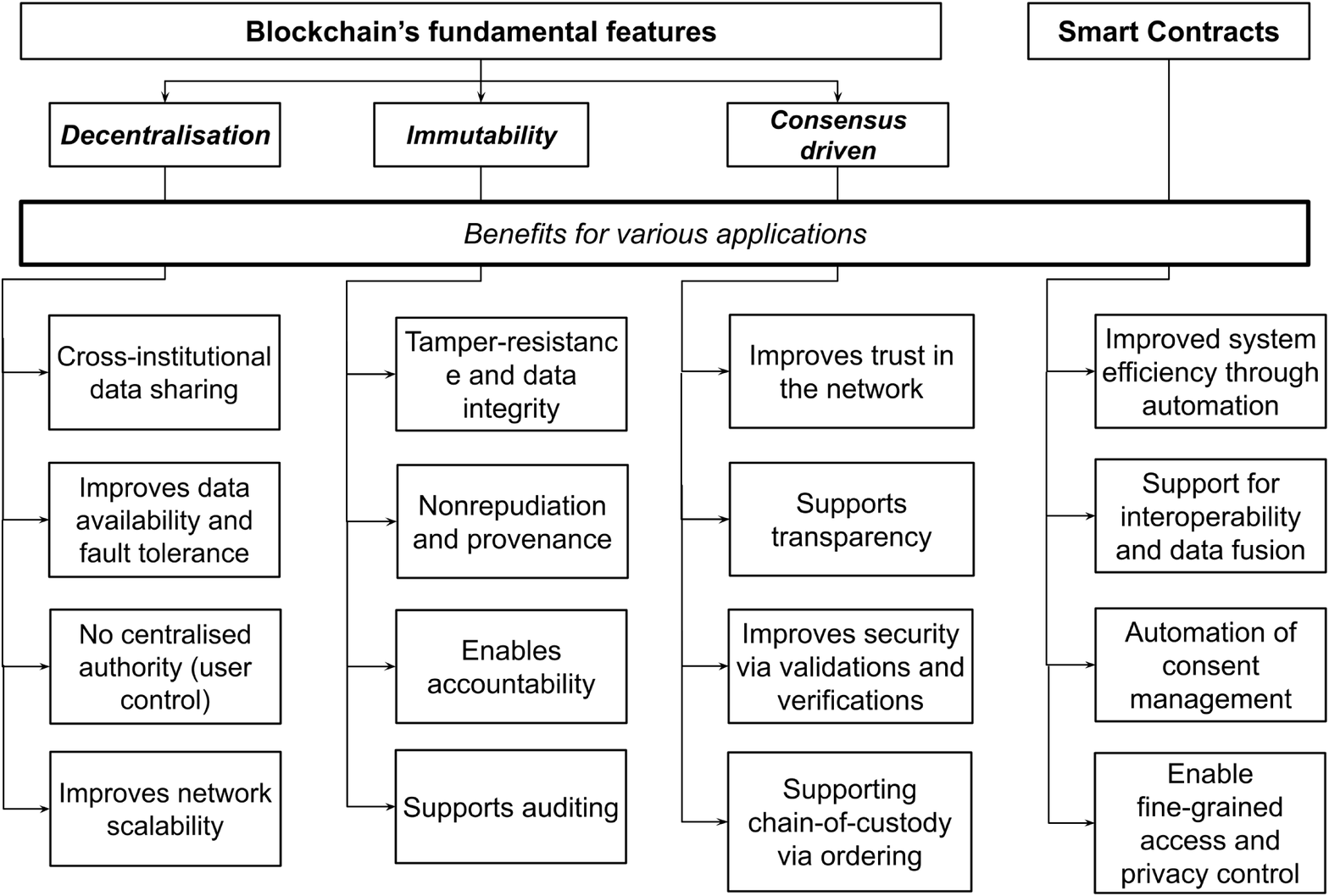}
\caption{Benefits by the usage of BC technologies and SCs}
  \label{fig1}
\end{figure}
\par Besides using SCs to eliminate the requirement of a trusted third party in multiparty interactions, there are other benefits that an SC provides. These include data fusion, consent management, fine-grained access control, and reduced bureaucracy and expenses. Moreover, blockchain, with its features such as tamper-resistance, decentralization, and transparency, provides a much-needed platform for secure deployment and execution of SCs. Therefore, in recent years, there has been a rapid increase in the popularity of SCs~\cite{Clack2016SmartCT}. Nowadays, SCs are being used in various BC-Apps to implement complex business logic. Same as the other software programs, SCs may contain bugs that could (and have been) lead to potential vulnerabilities that could be exploited with malicious intent. Since BC technology along with the SCs is mainly used in either financial applications (e.g., banking, and insurance, and trade of goods or services) or data-sensitive applications (e.g., Healthcare, and smart-grids~\cite{9103603}), these bugs can be exploited by malicious entities for financial gains or leaking sensitive data. Thus, it is vital to perform rigorous testing to ensure the development of bug-free SCs. Further, unlike other software, it is difficult to update an SC once it is deployed. Therefore, it is critical to verify and validate SCs before their deployment to avoid serious adverse consequences. Figure~\ref{fig1}, shows the key benefits that BC and SCs bring while being used in different application areas. These inherent features are the reason behind the rapid increase in the usage of these technologies in various domains. Still, at the same time, as these technologies are new in the market and thus are immature, there is a need for rigorous testing solutions to ensure their correct usage in any real-world application.

\subsection{Blockchain-based applications}

Although the BC has first emerged as a key underlying technology in the Bitcoin system, its unique features have attracted a huge number of other applications (financial as well as non-financial) that want to explore its usage for various benefits (e.g., data integrity, enhanced security, transparency, and data sharing). The organizations are exploring different options that are secure and robust and can share the information in a transparent way to provide an environment that can be trusted by the end-users. BC supports such transparency through decentralization. However, deploying BC as a solution could also have specific challenges, such as low scalability, high energy-consumption, integration problems, privacy, and security issues. To overcome these challenges, the BC-Apps must carry out essential testing measures. In particular, the BC technology used for an application does not work in isolation. It must interact with various components of the application infrastructure in which it is being deployed. Therefore, when a BC-based solution is being designed for any target application, the developers need to identify the essential interfaces required to ensure a secure and efficient interaction between the BC components and application entities. In particular, the BC components need to be integrated with other components or systems of a target application that enables access to the real-world data and events. These integration points require the design of well-defined application programming interfaces (APIs). APIs can also allow other foundational technologies, apart from the BC technology, that can also be incorporated as part of an overall BC solution. 

\begin{figure}
\centering
  \includegraphics[scale = .23]{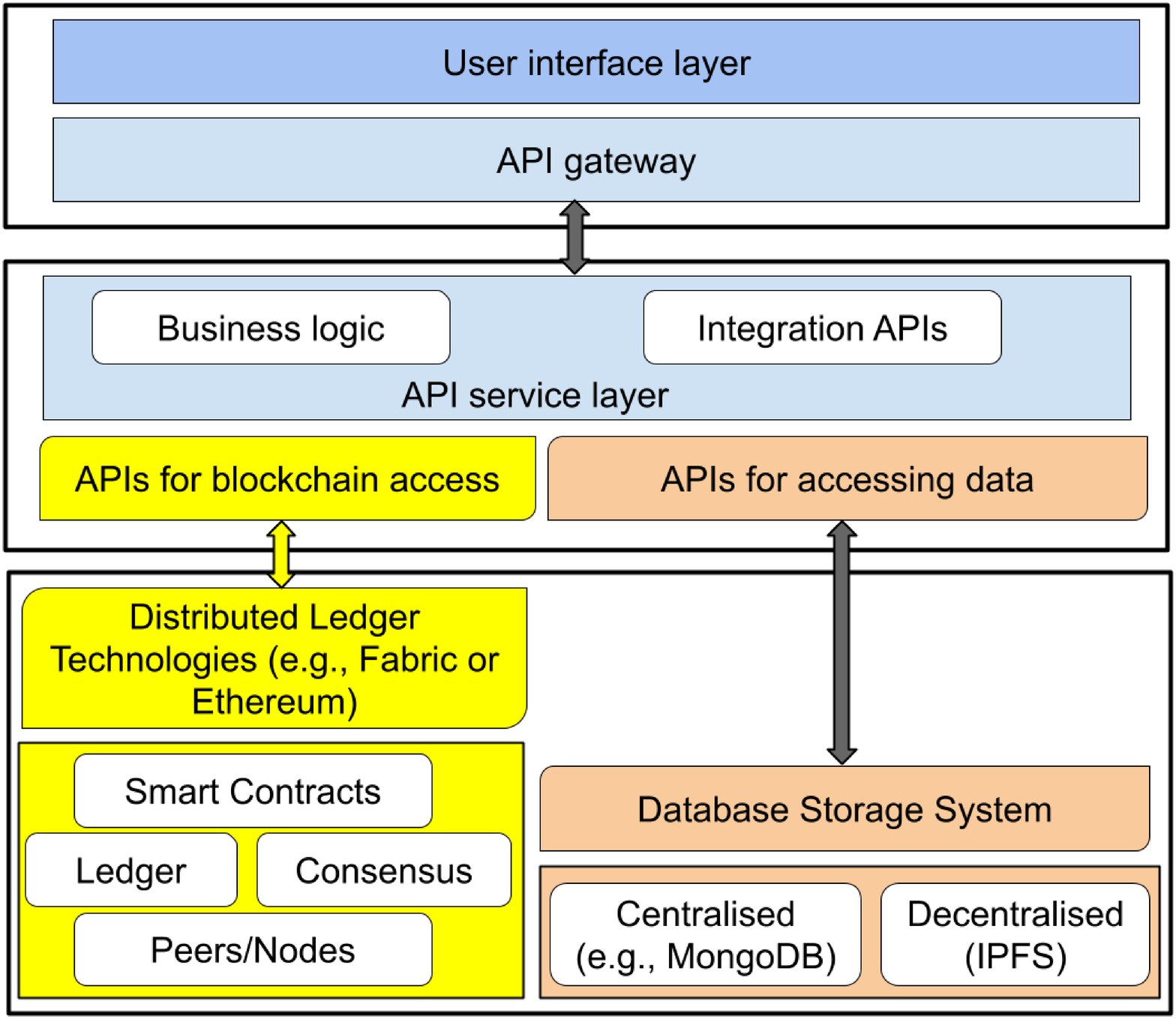}
\caption{Generic reference architecture for BC-Apps (shows the key components and their interactions that needs to be tested)}
  \label{fig2}
\end{figure}

\par In most of the existing BC-based application scenarios, the BC is considered as an add-on technology in existing business processes. Therefore, it necessitates testing tools to verify and validate all integration points between the BC and the application. However, over time a shift is seen towards the inclusion of BC from the start of the deployment phase of the applications rather than developing them in isolation followed by their integration. It is expected that there will be multiple interfaces with the application when BC is integrated with it. To ensure consistency with the existing application processes, it is important to understand and test these interfaces so there are no disconnect points left after the integration process. During testing, the developers should ensure that testers have access to the APIs used to communicate with existing business processes so they can use them to validate proper communication between legacy code and the BC implementations.

\par To better understand the interaction interfaces and components that require testing in BC-Apps, we provide a generic framework of BC-Apps in Figure~\ref{fig2}, and discuss it briefly by keeping our discussions within the context of testing requirements for such a BC-App. Figure~\ref{fig2} shows different components within the BC-based system and various interfaces that this system could have with the target application entities in which it is being deployed. Testers should identify the critical interfaces and components in BC-App that need to be tested to ensure the whole system's security and optimal performance. For instance, the sample framework in Figure~\ref{fig2} shows that there can be several middleware features that can be leveraged in a real-world BC implementation. These features include: (i) connectors to integrate different data sources (e.g., BC users, partner APIs, and cloud services), (ii) APIs for accessing and interacting with BC components, and (iii) connectors between BC and non-BC events in real-time, like machine learning, and analytics models. In particular, all the APIs need thorough testing for their envisioned functionalities. The goal is to ensure no functional issues exist, and all the service integrations are working as expected. Similarly, APIs used to access data storage systems should be tested for security (i.e., data protection from malicious or unauthorized entities), and performance (i.e., low-latency to data access operations).

\par Finally, as it is seen in Figure~\ref{fig2}, each BC-App implementation includes a set of core BC components such as SCs, peer nodes, consensus protocols, and storage systems (including both on-chain and off-chain) which are unique to a BC-App infrastructure. To ensure the BC projects' correctness, these individual components and their functioning should be tested thoroughly using different types (where applicable) of testing techniques. For instance, functional testing of SCs needs to be done to identify bugs and to check the implemented business logic's correctness. Security testing of SCs needs to detect vulnerability that might be exploited by a malicious entity after its deployment. Performance testing of the SCs could be done to check their execution and time complexity, leading to code optimizations. Since a large number of BC-based projects are implemented in data-driven applications, i.e., the data acts as an asset, and in the applications where data volume is huge (e.g., medical data sharing and managing digital forensics evidence), the off-chain data storage solutions are preferred due to performance and compliance reasons. Therefore, in such BC-App implementations, to ensure security (i.e., data protection) and good performance (i.e., low access latency), the data storage solutions and their interactions with other BC components and with application users should be performed. This survey mainly focuses on the testing techniques for the components and interfaces shown in yellow part in Figure~\ref{fig2}. Still, since performance and security testing are performed for the whole system, we discuss other needed components of Figure~\ref{fig2} as well.

\subsection{Testing types and techniques}
Software testing consists of a set of activities that aim to evaluate different software aspects to determine that the software meets its requirements. The software testing process consists of executing a software program and evaluating different software properties, with the overall intent of finding software errors. The testing process can only detect the existence of errors and can never confirm their absence. This is because exhaustive testing of non-trivial software programs is impossible, as the number of possible tests is practically infinite. Therefore, the primary objective of testing is to detect software errors. 

\par
\begin{figure}
\centering
  \includegraphics[scale = 0.19]{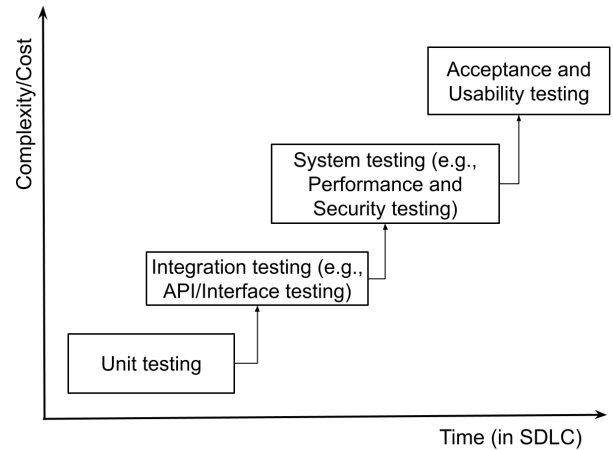}
\caption{Levels of software testing, depending on where it happens in SDLC}
  \label{fig_t_levels}
\end{figure}

 According to where testing happens in the software development life-cycle (SDLC), there are four major levels recognized: \textit{unit}, \textit{integration}, \textit{system}, and \textit{acceptance} testing. Unit testing aims to ensure that individual software code units are working correctly. Integration testing is performed to evaluate whether multiple units are working correctly when integrated. System testing analyzes the whole system's behavior, evaluating whether it is compliant with its specified requirements. Finally, acceptance testing tests whether the software works for its users. Both the complexity and cost of testing increases while moving from unit testing to acceptance testing, as illustrated in Figure~\ref{fig_t_levels}. This is because unit tests are more easily automated compared to other types of tests. 

\par Furthermore, there exist many different types of software testing.  At the most basic level, software testing can be \textit{static} or \textit{dynamic}. Static testing does not involve software execution, but performs the checking of source code structure, syntax and data-flow, and is thus better named static analysis. Typical approaches to static analysis include inspections, reviews, and walkthroughs. Static analysis can be used with \textit{formal verification} methods for proving the correctness of software. Formal verification normally requires a formal specification of software, which is used to provide a formal proof of the software correctness. Some approaches to formal verification include \textit{model checking}, \textit{deductive verification}, \textit{automated theorem proving}. Software verification can be contrasted to \textit{software validation}, which aims to check whether software satisfies its intended use (also associated with dynamic testing). 

\par Dynamic testing, on the other hand, involves software execution, while feeding inputs and producing outputs. Test cases are typically developed specifying test inputs and outputs, and the goal of testing is to check whether the actual outputs conform to the expected outputs. Dynamic testing techniques can be further classified as \textit{functional} and \textit{non-functional} testing. Functional software testing aims to evaluate whether software is compliant with specified software functional requirements. Functional testing techniques can be further classified as \textit{white-box} testing and \textit{black-box} testing. White-box testing aims to verify the internal structure of the software, and is usually performed at the level of unit testing. Specific white-box testing techniques include \textit{API testing} and \textit{mutation testing with fault injection}. Black-box testing aims to examine the functionality of software without having any knowledge about the software internal working. It answers the question of “what” the software does, not “how” it does it. Black-box testing is usually performed at the level of integration testing, system testing and user acceptance testing. Typical black-box testing techniques include \textit{model-based testing}, \textit{fuzz testing}, \textit{pairwise testing}, \textit{boundary value analysis}, \textit{equivalence partitioning}, and \textit{exploratory testing}. 

\par
\begin{figure}
\centering
  \includegraphics[scale = 0.16]{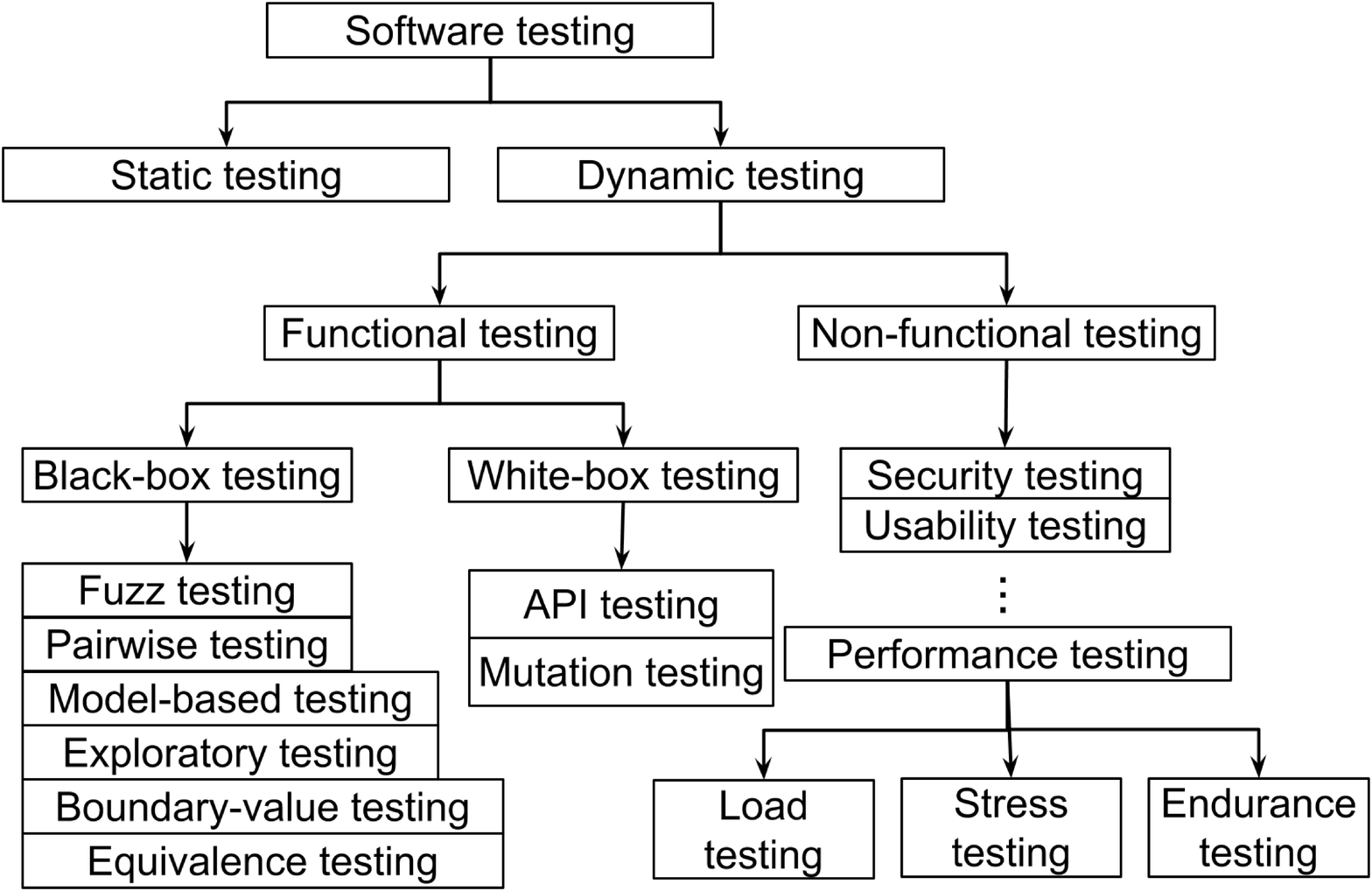}
\caption{Types of software testing}
  \label{fig_t_types}
\end{figure}

 Non-functional testing aims to examine non-functional requirements of software, for example how does the software perform under unforeseen (at design time) circumstances, or how does the software recover from failures. Some types of non-functional testing techniques include \textit{security} testing, \textit{performance} testing or \textit{usability} testing. Security testing aims at detecting threats, vulnerabilities and risks within software, to prevent attacks from intruders. Performance testing assesses how the software responds in conditions of a given workload. Typical types of performance testing include \textit{load} testing, performed to assess the response of software under a given load, \textit{stress} testing, performed to assess the response of software under upper limits of capacity, and \textit{endurance} testing, performed to assess software response under continuous load. Usability testing checks how easily the software can be used by its users, with the overall goal to improve user experience. Different types of software testing are illustrated in Figure~\ref{fig_t_types}. The presented classification of software testing types is not exhaustive, but informative, introducing the main concepts and types of testing. For the detailed description of the software testing field, we point interested readers to other research works, such as~\cite{beizer2003software} and ~\cite{pezze2008software}.

\section{Challenges in testing of BC-based applications}
This section presents the various challenges inherent to testing the BC-App implementations. Table~\ref{T:1} provides a list of challenges along with their description and the key reason behind the existence of the challenge. The identified key challenges are as follows.

\begin{table*}
\caption {Challenges in BC-App testing}
\centering
\scalebox{1.2}{
\begin{tabular}{|P{6cm}|P{8cm}|}\hline
\textbf{Challenge} & \textbf{Description and Why does it exist}\\ \hline
\multirow{3}{*} {Lack of best practices in developing BC-Apps} & heterogeneous domains (technical, non-technical, or legal) knowledge is needed \\  \cline{2-2} &  limited technological understanding \\ \cline{2-2} & conceptual and architectural ambiguity  \\\hline

\multirow{2}{*} {Lack of specialised tools} & lack of debugging/testing tools for the new programming languages used for SCs \\ \cline{2-2} & lack of automation tools for BC network deployment  \\\hline
\multirow{3}{*} {Lack of best practices to design test strategies} & lack of guiding procedures in designing and developing BC-Apps \\ \cline{2-2} & BC ecosystem is large and complicated \\ \cline{2-2} & BC-App ecosystems are highly complex and heterogeneous  \\\hline
\multirow{3}{*} {Blockchain immutability} & errors from BC-App users can result in irreversible losses as transactions are irreversible and immutable \\ \cline{2-2} & forces strict requirements for validation and verification of SCs before their deployment on BC making SC testing highly critical \\ \cline{2-2} & right-to-be-forgotten might be a lawful requirement in some BC-Apps \\\hline

\multirow{3}{*} {Performance evaluation} & 
generating real-world workloads and faultloads is challenging \\ \cline{2-2} & BC ecosystem is complex and it includes components with high randomness \\ \cline{2-2} & design and deployment of a SUT with high fidelity to production environment is difficult, time consuming, and expensive
  \\\hline
\multirow{3}{*} {Dependability assessment} & distributed nature of BC ecosystem components, techniques, and operations \\ \cline{2-2} & lack of automated dependability assessment tools \\ \cline{2-2} & requires a clear separation between system level, network level, and user level faultloads  \\\hline

\end{tabular}}
\label{T:1}
\end{table*}

\begin{itemize}
\item \textbf{Lack of best practices in developing BC-Apps:} It includes the lack of technology understanding, and the lack of skills or experience in designing and developing BC-Apps. Moreover, the lack of standardization in the usage of concepts and terminologies (i.e., conceptual and architectural ambiguity) in the BC ecosystem, which consists of BC software and technologies, leads to decrease in clarity, quality, and productivity during the BC-App development. To support the sustainable development of BC-Apps and enhance the software quality, specific best practices to improve the synergy between the system and the community working towards open source contributions in developing test-suits would be highly beneficial. Comparatively, BC is a new technology, and to adopt it in an application (e.g., healthcare or supply-chain management), one requires an in-depth understanding of its concepts and potential. Although it was introduced in 2008 by Satoshi Nakamoto as a key underlying technology in the Bitcoin system, it is still far from its potential adoption due to the lack of above mentioned best practices. Moreover, technical, non-technical, and legal (e.g., GDPR compliance rules) expertise, along with domain knowledge, are critical for effective and exhaustive testing of BC-Apps. Specifically, due to its business-critical nature, finance and legal professionals show high interest in BC-oriented software engineering. At the same time, tech boot camps for BC developers are blooming. \newline
Even with the awareness of this challenge, addressing it is not an easy task because acquiring the required additional skills (e.g., familiarity with new programming languages such as Go and Solidity) or understanding the best practices to implement BC-Apps are expensive. Furthermore, even with a successful implementation of BC-Apps, their testing is a highly specialized area that needs experience (i.e., proven expertise) and a rigorous testing approach. Lacking a strong ability to conceptualize, standardize, and abstract complex concepts underlying BC might lead to many BC technology issues. Furthermore, there is a need to open up new professional roles absent in conventional software development. In particular, the BC sector will need professionals with a well-defined skills portfolio comprising expertise such as finance, law, and technology. For instance, a new role could be an intermediary between focused business contractors with low technology expertise and IT professionals.

\item \textbf{Lack of specialized tools:} 
To ensure that the testing is done correctly, tools are needed to implement a system under test (SUT) which reflects the production environment with high fidelity. The deployment of the BC network (i.e., SUT) for testing a BC-App requires huge effort, and there is a lack of automation support for BC network deployment. Although a few deployment automation utilities are available in the market that offer such facilities via blockchain-as-a-service (BaaS), these BaaS are limited concerning the BC platform and hardware infrastructure that they support. Both, the availability and utilization of a SUT platform replicating the target implementation of BC-App are indispensable. If such a SUT is not available, then a significant part of time and resources need to be invested in setting up or spawning from the real implementation. To address this issue, few tools with open-source implementations exist that generate test cases which do not exactly reflect the test cases of a real-world scenario, but they can be still effective to test some of the advanced transaction functionalities. Moreover, compared to the public BC, setting up the test environment is easy in private BC, as it can be set up by configuring the deployment tools with customized functionality. However, the lack of proper development tools is a significant hurdle for BC implementations. In particular, a BC-App needs an integrated development environment that offers the required linters and plugins, a build tool and compiler, a deployment tool, a testing framework, and debugging and logging tools. Although few versions of these tools exist, they are not yet adequate to satisfy the needs of BC-App developers. \newline
Like any software project, a BC-App project also needs tools to perform various sorts of testing to test its components and functionalities. Therefore, if testers do not have a proper setup and the right tools, they will likely fail to provide comprehensive testing for the target BC-based system. Since BC technology is still evolving, the testing toolset is limited in nature, the available tools are not standardized, or they include a limited set of testing features. However, various open-source tools, such as Ganache, Hyperledger composer and caliper, Ethereum Tester, Exonum Testkit, and Embark, can be used for BC-Apps testing. The choice of tool(s) heavily depends on the target application and BC platform it uses. As a BC-App consists of various components of different types, such as SCs, peer-to-peer networks, distributed systems, and consensus protocols, it will need a set of specialized testing tools to test each of these components. For instance, several programming languages like Go, Solidity, and Ruby are increasingly used in BC-based projects. Since programming bugs are universal in software application systems~\cite{4407730}~\cite{5928349}, it advances the requirement for enhanced testing and debugging suites that are distinct to these languages, tailored upon the most popular BC languages. Undoubtedly, Java testing tools (i.e., provision for testing Java code via libraries and built-in tools) have experienced much more testing than Go lang. Also, as BC-App projects work with the BC, which is distributed by definition, testing in insulation would need tools that suitably produce mocking objects proficiently simulating the BC.

\item \textbf{Lack of guiding procedures to design test strategies:} 
The rapid increase in the popularity and demand of BC technology usage in several different applications may have pushed the BC developers to provide BC-based solutions in haste. This implies that testing is given less importance over programming and development process, leading to the BC-App development ecosystems with few or no dedicated tester(s) to investigate and evaluate the developed product. In particular, the testing strategy that is in use for developing BC-based solutions is immature. Thus, it leads to inefficient testing of a BC-App, which is either tested improperly (e.g., testing a module repeatedly) or not at all. Moreover, the BC ecosystem's complexity can dynamically expand the scope of minimal set of tests (i.e., sanity check) to a significant margin. Therefore, carefully (i.e., dedicating enough time and effort) preparing a testing strategy is vital in creating applications with BC technology. In particular, the more well-thought-through the testing strategy is, the higher the possibility of developing a quality product that performs at its best. \newline
The main factor determining the expected level of validation and testing required for a BC-App project depends on the BC-platform choice for its implementation. Suppose it uses a public platform such as Etherum or Openchain. In that case, the efforts will be comparatively less than a self-setup/customized platform that is purpose-built for an organization's needs. Such a BC-App implementation needs more setup and effort in testing. It is because the popular BC platforms have recommendations and guidelines on the level of tests required. Also, they have more mature methods developed over time, and these methods have been adopted by a vast community that further helped improve them. In contrast, an in-house BC-App implementation needs a detailed test strategy framework based on the functionality that is customized or developed.

\item \textbf{Blockchain immutability:} 
Implementing BC-based systems without paying special care to immutability carries a significant asset risk to institutions or users, particularly in financial applications. This is mainly because BC transactions are irreversible, and not having adequate controls to avoid redundancy and provide additional safety is a considerable challenge seen in the BC domain. Often, the transactions are used to invoke various functions in SCs. Therefore, the users must be aware of the possible output(s) of a transaction that they are performing. The SC developers could also place the required checks that invalidate a transaction if they try to invoke a wrong function in the SCs. To this end, testers need to ensure that SCs are checked for such possible transaction validation and verification codes before their deployment in the BC network. \newline
Immutability is an inherent property of BC that ensures its data integrity and auditability. Moreover, it supports BC's secure and transparent nature. Immutability guarantees that the data stored on the BC ledger are tamper-proof (i.e., it cannot be removed or modified). Although this property is highly desirable for several applications and is one of the key reasons that attracted a lot of attention towards BC technology, it repudiates many privacy requirements and data protection rights when personal data is involved as a BC asset. Among others, it disputes the Right-to-be-Forgotten (RtbF) described in the new EU data protection act (i.e., GDPR), according to which individuals have the right-to-delete their data if specific provisions apply~\cite{5928222}. Therefore, testers need to have adequate knowledge about the various data protection acts, and compliance testing needs to be carried out to ensure that the proposed BC implementation does not breach any of these acts. This implies that immutability calls for an interdisciplinary approach to testing. 

\item \textbf{Performance evaluation:} 
The aim of performance testing in BC typically includes identifying the bottlenecks, defining the metrics (e.g., latency and transactions per second) for tuning the system, and assessing whether the implementation design is production-ready. Accurate and comprehensive performance testing with varying workloads and faultloads is a key to gather insights into how the BC-App will perform in both production environments and under specific loads and network conditions. In a BC-App, there exist many components, and the overall performance of the BC implementation is tightly coupled with the individual and combined performance of these components. This, along with several dynamic events (e.g., rate of input transactions, network conditions, and dependencies), makes the end-to-end performance testing a significant challenge in the BC ecosystem. For instance, testers need to predict variances in their performance test-cases because transaction commit latency varies with the size of the P2P network and the transactions' volume. Data types and server locations can also influence it. Moreover, for performance testing, a high-fidelity replica of the production scenario is needed, but replicating the real-world transactions and the transaction processing latency is difficult. Let’s take an example to understand this issue better. To initiate a transaction in the Bitcoin system, miners have to confirm and validate the transaction, which could get delayed due to a surge in usage. Also, the distributed ledger that powers the BC requires to reflect the same order of transactions at each network node. Since the latencies across different consensus mechanisms might vary, testers have to perform peer/node testing to ensure the consistency and performance of newly committed transactions. Furthermore, to ensure the integrity of the network and the ledger, all these newly added transactions should be provided in the proper sequence. Hence, considering the above issues, it can be challenging to replicate the production system in a dummy environment.  \newline
One way to overcome some of the above-mentioned performance testing challenges is to have all the details, such as block size based communication latency, network size, expected size of the transaction, and time taken to process different queries stored on ledger data. In particular, testing for block size and ledger size is essential, as improper validation of these elements leads to BC application failure. Moreover, automated performance testing could be a key to assessing the overall scalability of the BC ecosystem. Finally, in BC-Apps, identifying the adequate trade-off between the consistency of ledger data and the availability and partition tolerance by setting the optimal values of different network parameters during performance testing is critical to the success of the target BC-Apps.

\item \textbf{Dependability assessment:} 
During the performance testing, one of the critical assessments that need to be performed by using a comprehensive set of faultloads is the SUT's dependability. However, due to the distributed nature of several components, techniques, and operations, performing dependability assessment is a challenging task in BC-Apps. Moreover, the rapid increase in BC-Apps usage in several critical application domains requires proper means to test and validate BC-App’s fault-tolerance levels and performability. At present, such assessments are partially conducted because the developers blindly trust the chosen BC platform. Although, proper assessments are needed not only to have evidence of the offered guarantees (e.g., performance and security levels), but also to select a BC platform over another based on the tested support for dependability assessment. Fault-injection, i.e., emulating custom faults in the SUT, is considered one of the most reliable techniques to validate software or distributed systems' dependability. In the fault-injection process, faultloads, with different granularities, are sent at different levels in the target systems to exercise the SUT under a various and rich set of possible issues (e.g., vulnerabilities and faults) that might occur in practice. Therefore, the need for developing the tools that can produce the required type of faultloads at different levels (i.e., system, network, and software) is a key challenge for efficient dependability assessment. For instance, at the system level, the faultloads should test the ability to tolerate process hangs and memory leaks. At the network level, the faultloads should test the ability to handle network partitions and message losses. Finally, to automate the dependability assessment procedure while following an experimental test-case, the evaluation tools must include an experiment orchestrator, execute the system with a workload, and stimulate the faultload at one or more levels.
\end{itemize}

\section{State-of-the-art on testing for BC-based applications}
\label{sota}
\begin{figure}
\centering
  \includegraphics[scale = .26]{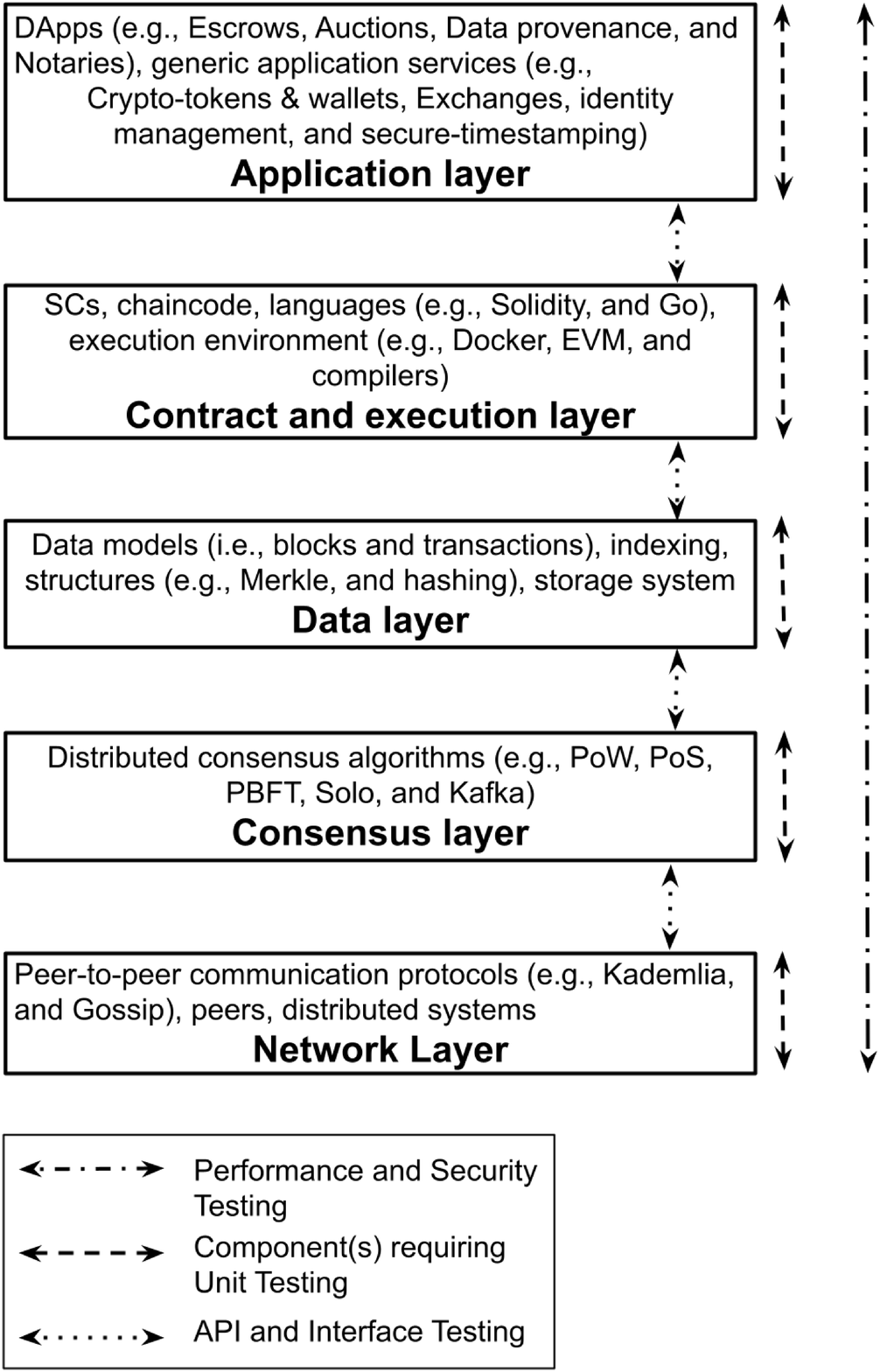}
\caption{Abstraction layers in a typical BC-App architecture}
  \label{stack}
\end{figure}

This section presents a detailed discussion on the state-of-the-art efforts towards testing vital components and functionalities that are being used in different BC-Apps. In this paper, we do not aim to provide a survey on functional and non-functional testing techniques of software systems that are not based on BC technologies, as there already exist many such studies~\cite{7123673}~\cite{5487526}~\cite{1122551}~\cite{3020266}. Figure~\ref{stack} shows a layered architecture for a typical BC-App. Here each layer includes a set of components, and the layers may or may not communicate with their adjacent layers. The figure also shows that different types of testing could span over one or components residing in multiple layers. In this work, we only cover the testing-related efforts that are new to the BC technology and those that might arise in different applications when integrated with BC. In particular, first, we start with the survey of existing tools and techniques used for discovering bugs and vulnerabilities in SCs. SCs implement the business logic of a target application. Therefore, they are considered one of the key components in BC-Apps, and thus need significant testing efforts. The same can be concluded by looking into the huge amount of research work done in this direction. Second, we discuss the research works that focus on providing techniques to perform performance testing of BC-Apps.  Testing the performance of the implemented BC-Apps is paramount to determine if they are ready for production. Third, we present the research efforts that identify or provide solutions for various security threats in a BC implementation. Specifically, we cover the research works that perform testing for issues such as confidentiality, data integrity, authentication, access control, authorization, and non-repudiation. Next, we provide a discussion on the existing testing solutions for different BC networking infrastructure components. These components include peers, consensus algorithms, and peer-to-peer networks. Finally, we present a discussion on research efforts that provide testing solutions for APIs and integration interfaces in the scope of BC-Apps.  

Please note that in this study, for the sake of simplicity, we refer to the terms SC testing and API/Interface testing as if they were performed on a BC component level (i.e., unit testing). Our study of SC testing, however, includes the testing efforts on SC security and performance as well. Further, we use the terms Performance and Security testing covering the testing across the whole BC-App stack (see Figure~\ref{stack}). 

\subsection{SC testing}
\label{SC-test}
The presence of suboptimal coding and bugs or vulnerabilities (e.g., buffer overflows, command injection, and cross-site scripting) in SCs after their deployments on BC network causes the following two major problems: (i) wastage of resources (e.g., gas and computing hardware) and delay in transaction processing due to presence of suboptimal code in SCs, and (ii) security threats caused by malicious entities by exploiting the vulnerabilities present in the SC. Sometimes, even simple bugs could open doors for new vulnerabilities in SCs. Both these problems could (and have) led to financial and non-financial (e.g., sensitive data) asset losses in BC-Apps. Moreover, such issues decrease the users' trust level, which could further harm the organization's reputation that is using the BC-Apps. Therefore, it is vital to equip developers and testing teams with tools and techniques to discover and fix suboptimal codes and vulnerabilities in SCs before deployment. To this end, the researchers have proposed several analysis approaches to discover and mitigate bugs in SCs. These approaches can be broadly classified into the following categories:
\begin{enumerate}
    \item Static analysis is a method of examining or analyzing a source code or compiled code without executing it. It investigates all potential code behaviors, vulnerable patterns, and flaws that can arise during the program's run-time.
    \item Dynamic analysis checks a program code at run-time or when it is being executed, acting  as an attacker that provides malicious code or anonymous inputs to the functions in a target program.
    \item Formal verification is a method to use theorem provers or formal techniques to prove the existence of specific properties (e.g., functional correctness, run-time safety, soundness, and reliability) in a program code.
\end{enumerate}
Next, we discuss all these three SC-testing approaches in detail by surveying the state-of-the-art. 

\begin{table*}[ht!]
\caption {SC vulnerability analysis and testing tools (Part 1)}
\centering
\scalebox{1.1}{
\begin{tabular}{|p{1.5cm}|p{3.3cm}|p{3.2cm}|p{3.3cm}|p{2cm}|}\hline

\textbf{Proposal} & \textbf{Brief description} & \textbf{Problem(s) addressed or Motivation} & \textbf{Analysis type and Methodology} & \textbf{BC platform and Dataset} \\ \hline

\cite{8823898}, (2019) &
Slither: a static analysis framework with features like speed, robustness, and adequate tradeoff between detection rate and false positives &
automated vulnerabilities and code optimization detection, assist users with code review &
static analysis, dataflow and taint tracking, Static Single Assignment (SSA) &
Ethereum, manually reviewed 1000 most used contracts \\ \hline
\cite{8445052}, (2018) &
SmartCheck: an extensible static analysis tool &
scalable and fast to find vulnerabilities by matching predefined patterns &
XML-based intermediate representation, XPath patterns &
Ethereum, 4600 verified contracts  \\ \hline 
\cite{Mueller2018}, (2018) &
Mythril: a security analysis tool, low scalability due to path explosion and unsolvable paths issues &
high accuracy due to exploration of  all execution paths &
static analysis, symbolic analysis, control flow checking &
Ethereum \\ \hline
\cite{43243780}, (2018) &
Securify: a static security analyzer, scalable, fully automated, and can prove SC behavior as safe or unsafe w.r.t a given property &
support for capturing critical violations, low false positive, and high code coverage &
dependency graph to obtain precise semantic knowledge, symbolic analysis, designated domain-specific language &
Ethereum, uses more than 18000 real-world SCs Etherscan \\ \hline
\cite{VIGLIANISI2020110647}, (2020) &
SoCRATES: highly configurable test-case generation framework, simulate composite interactions of users using federated society of bots &
locating programming bugs buried in complex and articulated interactions supported by the SCs, detect known errors, and detect previously unknown defects &
dynamic analysis using tests consist of different behaviour configurations, functional invariants check &
Ethereum, 1905 real SCs from Etherscan \\ \hline
\cite{8859497}  (2019), \cite{8719542}  (2018), \cite{3238177} (2018) &
ContractFuzzer: a fuzz testing service to test a set SCs as a whole &
new test oracles to precisely detect real-world vulnerabilities, addresses path explosion problem, low false positives &
dynamic analysis, static analysis, Application Binary Interface (ABI) analysis, fuzzing-based testing, instrumentation &
Ethereum, 6991 real SCs from Etherscan \\ \hline
\cite{8782988} (2018), \cite{8732934} (2019) &
Survey: detailed analysis report after evaluating various (27) Ethereum SCs testing tools by installing and testing them &
correctness (using formal methods) and security verification papers are studied &
n/a &
n/a \\ \hline
\cite{8945837} (2019) &
defines path-based test coverage criteria for systematic testing (complete transaction basic path set and bounded transaction interactions) of SCs &
shows ineffectiveness of existing tools in detecting logic errors and unknown vulnerabilities that do not match any patterns in SCs, detecting more number of faults than SOTA testing &
mutation for fault generation, fault detection via k-bounded transaction coverage, statement coverage, dynamic testing &
Ethereum, Pool-Shark application consists of 12 SCs \\ \hline
\cite{3386022} (2020) &
Behavioral simulation-based SC verification, automated verification of unannotated contracts against functional specifications &
large scale verification of SCs to detect vulnerabilities &
formal verification methods, loop invariants, contract invariants &
Ethereum \\ \hline
\cite{SINGH2020101654} (2020) &
Survey: formalization approaches for SC verification &
reviews Domain-Specific Languages (DSLs) and other formal languages, and tools and frameworks used for formalizing SCs &
n/a &
n/a \\ \hline
\cite{453360615}, (2019) &
NPChecker: detecting non-deterministic payment bugs in SCs &
provide tool that search bugs beyond the predefined vulnerability patterns &
static analysis, code instrumentation &
Ethereum, 30K online contracts (3,075 distinct) from mainnet \\ \hline
\cite{8946219}, (2019) &
Deviant: tool to automatically generate mutants of a given Solidity implementation to test SCs &
help developers to create high-quality SC source code, shows limitations of just using \textit{statement and branch coverage criteria} during testing & 
dynamic analysis, mutation testing with a combination of various mutation operators &
Ethereum, 3 projects with a total of 67 contracts \\ \hline

\end{tabular}}
\label{table:2}
\end{table*}

\begin{table*}[ht!]
\caption {SC vulnerability analysis and testing tools (Part 2)}
\centering
\scalebox{1.1}{
\begin{tabular}{|p{1.5cm}|p{3.3cm}|p{3.2cm}|p{3.3cm}|p{2cm}|}\hline

\textbf{Proposal} & \textbf{Brief description} & \textbf{Problem(s) addressed or Motivation} & \textbf{Analysis type and Methodology} & \textbf{BC platform and Dataset} \\ \hline

\cite{8813297} (2019) &
a clone detection approach for SC using SC birthmark (i.e., representation to abstract EVM bytecode and its business logics) &
vulnerability discovery and deployment optimization based on SC clone detection &
EVM bytecode, symbolic execution &
Ethereum, Mainnet Etherscan \\ \hline 
\cite{3404365} (2020) &
EShield: automated security enhancement tool for protecting SCs against reverse engineering (RE), shows protection against 3 such tools, extra gas cost needed &
traditional methods like obfuscation and encryption to protect code from RE does not work well with Ethereum SCs &
control flow graph, bytecode protection with anti-pattern insertion &
Ethereum \\ \hline
\cite{3385990}, (2020) &
Ethainter: analyzer to check information flow with data sanitization in SCs &
detects composite information flow violations, scalable, provide adequate tradeoff between precision and completeness &
static analysis, tainted information flow and guards, mutually-recursive datalog rules &
882000 from Ropsten testnet, and small random sample of contracts from Ethereum mainnet (240K)  \\ \hline
\cite{9054822}, (2020) &
ETHPLOIT: automated exploit generation tool for SCs, lightweight methods to solve the unsolvable constraints and BC effect's problem &
target the consequences of attacks instead of the code patterns, covers more exploits &
static taint analysis, iterative constraint, fuzzing-based exploit generation, path constraints &
Ethereum, 49,522 SCs from Etherscan \\ \hline
\cite{3293882333}, (2019) &
EthRacer: automatic analysis tool for SCs to detect event-ordering bugs, allows input as SCs before and after deployment on BC &
identify event-ordering (EO) bugs &
dynamic symbolic execution, combinatorial path, state explosion, partial-order reduction techniques, fuzzer &
Ethereum, 10 000 SCs from Solidity source code repository and Ethereum BC \\ \hline
\cite{3395363339}, (2020) &
SolidiFI: approach for automated and systematic evaluation of static analysis tools (six popular tools are considered for comparison) &
provides a systematic approach to evaluate the effectiveness of  static analysis tools in finding security bugs in SCs &
bug defects, bug injection, runtime activation of injected bug by external attacker &
Ethereum, bugs are injected into 50 SC’s source code \\ \hline
\cite{8949045}, (2019) &
ML-based detection of security vulnerability patterns &
existing ML-based approaches are not well-suited for SC  languages, static code analyzers alone can be highly time consuming &
static code analyzers for labeling, predictive model &
Ethereum, 1000 SCs \\ \hline

\cite{5099582}, (2020) &
MuSC: mutation-based testing tool to test SCs at large scale (generated 71,314 mutants for 2,393,500 transactions) &
focus on quality and completeness of the tests, novel killing condition to detect a deviation in gas consumption &
mutation-based testing, Abstract Syntax Tree (AST) &
Ethereum \\ \hline
\cite{3385412338}, (2020) &
Solythesis: a runtime automated validation tool for SCs to detect invariants violations & 
provides source-to-source Solidity compiler, provides instrumentation algorithm to minimize the number of BC state accesses, higher code coverage  &
invariant specification language, instrumentation optimizations, runtime validation technique &
Ethereum, 23 smart contracts from ERC20, ERC721, and ERC1202 standards \\ \hline

\cite{8945725}, (2020) &
SolAnalyser: automated security analysis tool, along with a tool that inject faults in SCs, compared with Oyente, Securify, Maian, SmartCheck and Mythril tools &
detects more number of vulnerabilities, scalable, and low false positives, large scale evaluation &
combination of static and dynamic analysis, ABI, mutation &
Ethereum, 1838 SCs, and 12866 mutated SCs \\ \hline
\cite{1078001}, 2020 &
SOLC-VERIFY: tool for source-level formal verification of SCs &
support to discover non-trivial bugs and prove correctness, allows modular verification, scalable &
formal verification, AST, Satisfiability modulo theories solvers &
Ethereum, 37531 SCs from Etherscan  \\ \hline

\end{tabular}}
\label{table:3}
\end{table*}

\begin{table*}[ht!]
\caption {SC vulnerability analysis and testing tools (Part 3)}
\centering
\scalebox{1.1}{
\begin{tabular}{|p{1.5cm}|p{3.3cm}|p{3.2cm}|p{3.3cm}|p{2cm}|}\hline

\textbf{Proposal} & \textbf{Brief description} & \textbf{Problem(s) addressed or Motivation} & \textbf{Analysis type and Methodology} & \textbf{BC platform and Dataset} \\ \hline

\cite{3404366}, (2020) &
Echidna: automatically generate tests to discover infractions in assertions and custom properties &
fuzzing based on custom user-defined properties to detect exploitable defects in SCs, easy-to-use, fast, and  high quality and code coverage &
fuzzing-based testing, property-based and coverage-driven fuzzing, assertion checking  &
VeriSmart and Tether BCs and their token contracts \\ \hline

\cite{3291303}, (2018) &
Survey/analysis: provides a comprehensive analysis of Oyente, Securify, Mythril, and SmartCheck tools &
reviewed the automated SC security testing tools concerning their vulnerability detection ability and accuracy scores &
static analysis, bytecode level, symbolic execution &
Ethereum, 10 SCs from Ethernaut, and Trail of Bits \\ \hline

\cite{8939256} (2019) &
SoliAudit: SC vulnerability assessment tool, uses solidity machine code as learning features, detects 13 types of vulnerabilities &
detects vulnerabilities without referring to any predefined patterns &
machine learning, gray-box fuzz testing &
Ethereum, 18k SCs, 14,383 training samples and 3,596 test samples \\ \hline
\cite{9054825} (2020) &
SMARTSHIELD: a bytecode rectification system that fixes three SCs security-related bugs automatically, make the resultant SC more gas-friendly and secure against common attacks &
provide solution for detection of bugs as well as code fix issues, shows scalability, correctness, and cost reduction &
semantic-preserving code transformation, Contract rectification, CFG,, abstract syntax tree &
Ethereum, 28,621 real-world buggy contracts \\ \hline

\cite{2976749297} (2016) &
Oyente: a tool for SC testing against security vulnerabilities &
support bulk analysis, one of the first attempts in this direction, discover many new classes of security bugs in Ethereum SCs &
CFG, static analysis, constraint solving, symbolic execution & 19366 SCs from Ethereum BC \\ \hline
\cite{8449446}, (2018) &
ReGaurd: a tool for identifying specific type (i.e., Reentrancy Bugs) of issues in SCs &
supports bulk analysis of SCs, allow inputs as bytecode as well as solidity code &
dynamic analysis, CFG &
Ethereum \\ \hline

 \cite{8987460} (2019) &
MPro: a scalable automated testing approach for SCs &
proposes a detection approach for \textit{depth-n} vulnerability, based on Mythril-Classic and Slither tools &
symbolic execution, data dependency analysis &
Ethereum, 100 real-world SCs \\ \hline
\cite{praitheeshan2020security}, (2020) &
Survey of security analysis approaches for vulnerabilities in SC  &
discusses security issues in SCs along with state-of-the-art analysis and detection tools/techniques &
16 ESC vulnerabilities, 8 vulnerability detection tools are analysed &
Ethereum \\ \hline
\cite{3370272337}, (2019) &
SolUnit: novel approach for fast SC’s unit testing through the reuse of SC deployment and test setup execution, 5 projects are evaluated using the proposed approach &
without breaking the principle of independent tests (i.e., cannot create a dependency between tests) &
source code analysis, safe/unsafe test methods &
Ethereum, number of SCs not specified \\ \hline
\cite{torres2020smart}, 2020 &
CONFUZZIUS: a tool based on hybrid fuzzing, compared with 5 STOA tools concerning code coverage and number of vulnerabilities detected  &
aim is to execute more code and find more bugs in SCs, addresses the issues related to environmental dependencies &
hybrid fuzzer, evolutionary fuzzing, constraint solving, data dependency, symbolic taint analysis to generate path constraints, mutation pool &
Ethereum, 27 real-world SCs collected from 17 GitHub repositories \\ \hline
\cite{yu2020SCRepair}, (2020) &
SCRepair: a tool for pre-deployment analysis and automated repair of bugs in SCs, uses Oyente and Slither tools for vulnerability detection &
searches among mutations of the buggy contract, uses parallel genetic repair algorithm to split search space, and to generate a patch for vulnerable SCs &
automated program repair, search-based, gas-aware, candidate patches, gas dominance relationship, genetic programming search &
Ethereum, 38225 SCs from Etherscan \\ \hline

\cite{9152791}, 2020 &
VerX: automated verifier to prove functional properties of SCs &
verifier to automatically prove custom and functional properties of SCs, fast and less-expensive approach &
formal verification, symbolic execution, delayed predicate abstraction &
Ethereum, 12 real-world projects (138 contracts), 83 safety properties  \\ \hline

\end{tabular}}
\label{table:4}
\end{table*}

\subsubsection{Static analysis methods}
The SC testing techniques which are based on static analysis approach can be further classified into several categories depending upon the methodology they use, such as symbolic execution (e.g., \textit{Oyente}~\cite{2976749297}, \textit{Securify}~\cite{43243780}, and \textit{Manticore}~\cite{8952204}), pattern matching~\cite{8445052}, control flow graph~\cite{7884650}, and Static Single Assignment (SSA)~\cite{8823898}. 
\par Authors in~\cite{2976749297} propose \textit{Oyente}, the first symbolic execution tool for Ethereum SCs that automatically detects popular vulnerabilities like transaction order dependency and reentrancy. It directly works with Ethereum Virtual Machine (EVM) bytecode without needing to access the high-level representations (e.g., Solidity, Serpent). It was evaluated on 19366 SCs extracted from the first 1,460,000 blocks in the Ethereum network, and it detected that 8,833 SCs potentially have the documented bugs. However, \textit{Oyente} only aims to detect potentially vulnerable contracts, leaving the full scale false positive detection as future work. Authors in~\cite{13274737} propose \textit{Osiris}, which is based on \textit{Oyente} with a specific focus on detecting integer related bugs like integer overflows. Moreover, it combines taint analysis with symbolic execution to reduce the number of false positives. Similarly, authors in~\cite{7884222} propose \textit{Mythril}, a tool to perform security analysis of Ethereum SCs by using concolic and taint analysis and control-flow checking. 

\par Apart from symbolic execution methods, there are works based on static analysis that use other methods. For example, in~\cite{Kalra2018ZEUSAS} an automated SC verification framework called \textit{ZEUS} is proposed, which uses abstract interpretation and model checking, and accepts user-provided policies. First, it inserts policy predicates as assert statements in the SC code, then it translates the code into an intermediate low-level virtual machine (LLVM) representation, and finally, it calls its verification procedure to ascertain assertion violations. Another tool called \textit{Securify} is proposed in~\cite{13243780}, a tool that first uses static analysis to extract semantic information about the SC bytecode using a dependency graph of the SC, and then safety pattern violations are checked. It also allows adding new patterns created using a designated domain-specific language, thus adds flexibility. \textit{SmartCheck}~\cite{8445052}, a static analysis tool that first changes Solidity code into an XML-based intermediate representation and then examines it against XPath patterns to find potential security, operative, operational, and development bugs. Also, \textit{Gasper}~\cite{7884650} and \textit{GasReduce}~\cite{13509255} uses static analysis to identify potential code optimizations in SCs at high-level and at bytecode instruction level by mainly focusing on dead code and loop optimizations. Other popular static analysis tools are \textit{Vandal}~\cite{brent2018vandal} and \textit{EtherTrust}~\cite{Grishchenko2018}.

\par Authors in~\cite{8823898} propose \textit{Slither} which uses static analysis methods to provide rich information, real-world contracts, speed, robustness, and an adequate tradeoff between detection rate and false positives. It provides automated vulnerability detection and code optimization detection. It also supports users to improve their understanding of the SC code by assisting them with code review, thus flagging bad coding practices. Moreover, it uses program analysis techniques such as dataflow and taint tracking to extract and refine the information, and it performs lexical and syntactical analysis on Solidity code. Authors in~\cite{453360615} propose \textit{NPChecker}, a tool that uses static analysis along with code instrumentation for detecting non-deterministic payment bugs in SCs. The novelty lies in the fact that it provides support to search bugs beyond the predefined vulnerability patterns.

\par Recently, authors in~\cite{3385990} provided a security analyzer for SCs, namely \textit{Ethainter}. It checks information flow with data sanitization in SCs to identify compound attacks that include an intensification of tainted information through many transactions, leading to extreme violations. The Ethainter implementation is highly scalable, shown by applying it to the entire set of unique SCs (around 38MLoC) on the Ethereum BC in 6 hours. The evaluation results show that using automatic exploit generation (e.g., killing about 800 SCs on the Ropsten network) and manual review, \textit{Ethainter} obtains high accuracy of 82.5\% real warnings for end-to-end vulnerabilities. Moreover, Ethainter’s adequate tradeoff between code coverage and false positives significantly improves other tools like Securify, Securify2, and teEther. Finally, authors in~\cite{3395363339} propose \textit{SolidiFI}, an automated tool that uses a systematic approach to evaluate existing static analysis tools to detect SC vulnerabilities. First, SolidiFI allows bug injection (i.e., code defects) at different SC locations to introduce predefined security vulnerabilities. Second, it analyzes the resulting buggy SC by using the static analysis tools and collects the bugs that remain undetected (along with the false positives) by these tools. SolidiFI evaluates six static analysis tools: SmartCheck, Manticore, Oyente, Securify, Mythril, and Slither, using 50 SCs injected with 9369 distinct bugs. The evaluation results show that several bug instances remain undetected by these tools despite their claims of detecting such bugs. Moreover, all the tools list several false positives. 

\subsubsection{Dynamic analysis methods}
Since dynamic analysis checks a programming application at runtime, it can replicate an attacker looking for vulnerabilities in a piece of code under testing. It does this by feeding malicious or anonymous inputs to the specific functions in the SCs. Researchers showed that static analysis tools detect few vulnerabilities as false negatives, but dynamic analysis tools can successfully detect them. Moreover, dynamic analysis methods can also validate the findings of a  static code analyzer. Next, we discuss some popular dynamic analysis methods along with some hybrid approaches that use both dynamic and static analysis techniques.

\par In~\cite{3274743}, a tool called \textit{MAIAN} is proposed that uses inter-procedural symbolic analysis and concrete validation to identify and verify vulnerabilities on trace properties (e.g., identifying SCs that endlessly lock funds or leak them to random users) of Ethereum SCs during runtime. MAIAN labels the malicious SCs in three categories, namely greedy (i.e., lock funds indefinitely), prodigal (i.e., releases funds to arbitrary accounts instead of legitimate owners), and suicidal (a random account kills the SC or forcibly execute the \textit{suicide} code). Authors in~\cite{8946219} present \textit{Deviant}, a mutation-based security testing tool for Solidity SCs. Deviant automatically produces mutants\footnote{Changes in software code expected to induce errors.} for a target test project and runs them against the predefined test-cases to assess their effectiveness. To reproduce several faults in Solidity SCs, Deviant presents mutation operators (different from the traditional programming constructs) for all the distinct features of Solidity according to its fault model. The simulation results acquired by running Deviant to test the three Solidity-based projects show that these tests have not yet achieved large mutation scores. It further shows that a test suite competent for the coverage statement and branch criteria of Solidity SCs does not surely give a high-level assurance of code quality. Such measurements advise Solidity developers to implement important guidelines that provide more effective test sets to deliver trustworthy code and decrease security risks. 

\par Authors in~\cite{3293882333} examine a family of bugs called event-ordering (EO) in BC-based SCs. These bugs are linked to SC events' dynamic ordering, i.e., its function calls, and could facilitate possible exploits of millions of dollars' worth of cryptocurrency. The paper proposes the use of a concurrent program analysis technique to formulate a generic class of EO bugs that arises due to long permutations of such events. The authors show that the technical challenge in detecting EO bugs in SCs (even in simple ones) is the intrinsic combinatorial dispute in the path and state-space analysis. Moreover, the paper introduces techniques based on partial-order reduction, which automatically extracts happens-before relations and numerous dynamic symbolic execution optimizations. Finally, an automatic security analysis tool called \textit{EthRacer}  is proposed, which runs on top of Ethereum bytecode and requires no input from users. The experimental outcome shows that most SCs do not manifest varieties in outputs. EthRacer analyzes 1000 SCs, out of which it flags 8\% for vulnerabilities. Moreover, when SCs do show distinct outcomes upon reordering, it is observed that they are likely to have an unintended behavior most of the time. The comparsion with Oyente reveals that EthRacer discovers all 78 true EO bugs that Oyente detects along 596 bugs that Oyente fails to find. Authors in~\cite{torres2020smart} propose a hybrid fuzzer called \textit{CONFUZZIUS}. It aims to provide higher code coverage and detect more bugs using the evolutionary fuzzing and constraint solving method. In particular, evolutionary fuzzing is applied on the shallow parts of SC, while constraint solving generates inputs that meet difficult conditions that restrict the evolutionary fuzzing from investigating deeper paths. Moreover, data dependency analysis efficiently generates sequences of transactions to create specific SC states that may hide bugs. CONFUZZIUS uses a more efficient fuzzing approach than ETHRACER because instead of using an entirely random transaction order, it uses read-after-write data dependencies between transactions. It is thus generating quicker and more useful combinations of transaction order dependencies.

\par In the scope of the dynamic analysis approaches, the researchers heavily use fuzzing-based methods (mainly mutation-based techniques) for runtime testing of SCs. However, SC fuzzing presents some unique challenges that are unusual in traditional fuzzer development approach~\cite{3404366}. The challenges are as follows: (i) a considerable amount of engineering effort is needed to simulate the semantics of BC execution, (ii) simulate transaction sequence generation~\cite{4222570}, and (iii) finding SC inputs that produce disordered execution times is not an unfamiliar concern, as in traditional fuzzing~\cite{13213874}. In BCs like Ethereum, where a certain amount of gas is needed to execute a transaction, SC design inefficiency can be costly, and malicious inputs can lock SCs by making all transactions require more gas than needed. Therefore, delivering a quantitative output to put an upper bound on the gas usage is a critical fuzzer feature, besides more traditional correctness restraints. To address the issues mentioned above, authors in~\cite{3404366} propose a security analysis tool called \textit{Echidna}, which is easy to use and configurable, and support high code coverage, and produces results faster. \textit{Echidna} works in two steps. First, it leverages a static analysis framework called Slither~\cite{8823898} to compile SCs and check them for constants and functions that directly handles Ether (ETH). Second, the fuzzing operations start with an iterative method producing arbitrary transactions using the following (i) Application Binary Interface (ABI) given by the SC, (ii) critical constants defined in SC, and (iii) any previously collected sets of transactions from the corpus. A counterexample is automatically minimized upon a property violation to report the smallest and simplest transactions that trigger the failure. Optionally, Echidna can also provide a transaction set to maximize the coverage over all the SCs.

\par Authors in~\cite{3238177} propose \textit{ContractFuzzer}, an accurate and comprehensive fuzzing-based approach to detect seven types of Ethereum SC vulnerabilities. It contains an offline EVM instrumentation tool that performs instrumentation of EVM so that an online fuzzing tool can oversee the execution of SCs to get the required information for vulnerability analysis. It also provides a set of new test oracles that can accurately detect real-world vulnerabilities within SCs. The systematic fuzzing performed on 6991 real-world Ethereum SCs showed that \textit{ContractFuzzer} had identified at least 459 SCs vulnerabilities, including the DAO and Parity Wallet. The performance analysis also showed that it detects more types of bugs, but it has lower false positives than Oyente. Furthermore, authors in~\cite{VIGLIANISI2020110647} propose \textit{SoCRATES}, a extremely configurable and extensible framework to generate test-cases to test SCs. SoCRATES uses a federated organization of bots to mimic the complex interactions among many users, and these bots interact with the BC as per a defined set of composable behaviors. In particular, the use of society of bots in \textit{SoCRATES} allows triggering faults simulating the multi-user interactions that are difficult to produce by using one bot. The aim is to spot programming defects hidden in complex and articulated interactions backed by the SCs. Also, to expose known and unknown faults in SCs currently published in Ethereum.

\par Finally, authors in~\cite{8945725} propose \textit{SolAnalyser}, a fully automated approach that uses static and dynamic analysis for vulnerability detection over Solidity SCs. SolAnalyser supports the detection of eight vulnerability types that are not supported by state-of-the-art tools. Moreover, it can be easily extended to support other kinds of vulnerabilities. ~\cite{8945725} also contributes a fault seeding tool that injects different types of vulnerabilities in SCs. SolAnalyser is evaluated by experimenting with 1838 real SCs from which 12866 mutated SCs are produced by artificially seeding eight distinct vulnerability types. The results show that SolAnalyser can identify the seeded vulnerabilities. Furthermore, SolAnalyser outperforms five popular analysis tools (i.e., Oyente, Securify, Maian, SmartCheck, and Mythril) by detecting all eight vulnerability types and achieving a high precision/recall rate. 

\par Although the static and dynamic security analysis approaches provide adequate code coverage, it appears that they are not enough on their own. This is because these approaches are not designed for testing SC’s functional correctness. A complementary approach could be to use model-based testing (MBT)~\cite{stvr456} in which test automation is based on a model. Recently, authors in~\cite{13417939} propose ModCon, a model-based testing tool that uses an explicit abstract model of the target SCs to derive tests automatically. ModCon complements other static validation methods in rendering more flexible and reliable quality assurance solutions. Moreover, most of the static and dynamic security analysis tools used for testing SCs are designed and analyzed against public BCs (e.g., Ethereum). They might not be suitable for enterprise SC applications. ModCon shows its effectiveness specifically for enterprise SC applications based on private BC platforms. It allows SC developers to input their test model for the SC under test. Naturally, the efficiency of MBT depends on the input test model and fault model (or test hypotheses). An exciting research challenge could be to investigate SCs and their faults to determine suitable fault models or test assumptions.  

\subsubsection{Formal verification methods}
\label{formal-methods}
Apart from the above mentioned static, dynamic, and hybrid analysis approaches to detect vulnerabilities in SCs, another approach to detect bugs is using formal verification methods. One key difference compared to the other testing methods is that that formal verification methods are used in the design phase to check if the design itself is correct. The most used techniques to carry out the formal verification are \textit{theorem proving}~\cite{1097833} and \textit{model checking} (also called property checking)~\cite{SELIGMAN201523}. Theorem proving requires well-known axioms and basic inference rules which are used to derive every new theorem or lemma that are needed for the proof. Since it applies to all systems that can be expressed mathematically, theorem proving is considered a flexible verification method. Moreover, it can be interactive, automated, or a hybrid between the two. On the other hand, model checking uses specific software to verify if a system’s finite-state model works as per its formal specification and correctness properties. Model-checking software first takes input from a user, including the finite state model of the SUT and the set of formally specified properties that it should have. Second, it checks if all the states satisfy the specifications. Moreover, authors in studies like~\cite{8666514}, and~\cite{8595076} show the usage of formal verification methods for the correctness verification of BC consensus algorithms.

\par \cite{9152791} proposes the first automatic verifier called \textit{VERX}, which can prove the temporal safety properties of Ethereum SCs. VERX's design consists of combining the following three ideas: (i) reachability checking via reduction of temporal safety verification, (ii) calculation of particular symbolic states within a transaction by using an efficient symbolic execution engine, and (iii) delayed abstraction, which approximates symbolic states into abstract states at the end of transactions. Based on the evaluation results obtained by considering 12 real-world Ethereum projects, it is determined that VERX automatically proves 83 temporal safety properties of SCs, demonstrating its practicality.  Moreover, the experiment results conclude that VERX is a practical framework for testing custom functional properties of SCs. However, it is challenging to scale the verifying attempts to a large number of SCs. In particular, while particular specifications of SCs are mandatory to prove its customized functional properties~\cite{9152791}, general specifications applicable to large classes of SCs would facilitate verifying SCs in a group. Ideally, the specifications of given SC classes can be written once and reused to test each class's SC. Indeed, general specifications need to be adequately weak to be mapped with a class's functional properties. Furthermore, genuinely general specifications need to be free of any particular SC's state variables because other SCs belonging to the same class typically differ in name, type, and number. However, such general specifications are not well suited for existing verification tools such as VerX~\cite{9152791} and solc-verify~\cite{1078001}, which assume that input SCs are annotated with expressions that refer to state variables (e.g., pre-conditions and post-conditions). Therefore, it poses a scalability issue since deriving such annotations for each SC from class-wide general specifications can be a labor-intensive process. Recently, authors in~\cite{3386022} introduce an automated method to check unannotated SCs against specs ascribed to a few manually-annotated SCs. Specifically, a concept of behavioral clarification, which signifies the inheritance of functional properties and an automated approach to inductive proof by synthesizing simulation relations on the states of related SCs, is contributed.

\subsection{Performance testing}
\label{per-test}
This section identifies the current mainstream techniques, critical evaluation metrics, benchmark workloads for BC performance testing (PT), the significant performance bottlenecks identified in various BC-based system, and the main challenges in BC PT. The largest number of BC technology applications is in data-driven domains, and using them in settings where database technologies have established dominance is becoming a new trend. But before employing BC instead of traditional databases, one question to ask is to what extent BC can handle data processing workload. Therefore, application developers need to have access to frameworks that help evaluate BC's potential in meeting the target application’s needs and help them identify and improve on the performance bottlenecks. For this reason, PT is crucial in BC-App development. Since performance is usually measured in end-to-end scenarios, PT focuses on various parameters, including performance bottlenecks, production environment, network latency, transaction sequence at each node, transaction processing speed, and responses needed from smart contracts. In particular, the end-to-end PT will depend on the performance of various individual component at different layers (refer Figure~\ref{stack}) of the BC stack. For instance, a network is critical, and testers should deep dive into network layers to calculate metrics like transaction throughput and network latency, among many others. These critical metrics can be measured using a mix of tools like ELK Stack~\footnote{https://www.elastic.co/elastic-stack} and Jmeter~\footnote{https://jmeter.apache.org/}. Furthermore, the health of connected peers (CPU usage, memory utilization, and disk i/o) can be monitored through the use of metricbeat~\footnote{https://www.elastic.co/beats/metricbeat}. Also, as the compound testing can have multiple endpoints, one should consider an end-to-end scenario for performance, which can lead to an automated PT to enhance the BC ecosystem's scalability.
\par In the literature, there exist different approaches to conduct performance testing of BC-based systems. The most common of these are benchmarking, monitoring, experimental analysis, and simulation techniques. These approaches aim to help developers/users effectively evaluate different performance characteristics and identify bottlenecks accordingly, to improve performance. Next, we discuss these approaches in detail, along with their advantages and limitations, starting with the benchmarking approach. Moreover, in Table~\ref{T:5}, we summarize and compare different types of BC performance analysis benchmarking tools. As it is seen in Table~\ref{T:5}, there are several efforts from academia as well as from industry to develop benchmarking tools for performance analysis of BC platforms. All these tools are open-source. Caliper and DAGBench are from industry (i.e., IBM, and IOTA foundation), while the rest are from academia. Both Caliper and DAGBench are well documented and active (i.e., new versions are being updated continuously), and they are being actively used within the research community for performance evaluation of their proposed BC-App solutions. On the other hand, apart from the BBB tool, the rest are inactive at present.     

\begin{itemize} 
\item \textbf{BlockBench:} The first open-source benchmark tool called \textit{BlockBench}~\cite{1064033} is developed to perfrom performance evaluation of private BC-based systems. The metrics it measures include TPS, latency, scalability, and fault-tolerance. Authors in~\cite{1064033} use BlockBench to compare the performance of four major private BC platforms (i.e., Ethereum, Parity, Fabric, and Quorum). The authors also claim that BlockBench can be used for the evaluation of other private BCs by extending the workloads and BC adaptors. The evaluation of the four BCs is performed by first splitting BC functionalities into four concrete layers, and then each layer is evaluated against different workloads. The evaluation results show that (i) the consensus algorithm is the main bottleneck in HLF v0.6 and Ethereum, and (ii) the Ethereum execution engine is less efficient compared to Fabric. BlockBench uses two workloads, namely YCSB~\cite{1807152} and Smallbank~\cite{1620587}, to quantify the four performance metrics for the target BC platforms. The performance evaluation shows various performance deficiencies in the comparison of BC-based systems. These deficiencies depend on the design choices made by developers at different layers of the BC's software stack. Moreover, the results showed that current BCs are not well suited for large-scale data processing workloads. Some key limitations of BlockBench are as follows: (i) deployment of BCs to be tested is managed through bash scripts that do not offer abstractions over the targeted testbed, (ii) it collects metrics about performance (latency and throughput) but does not include system metrics like CPU, memory or disk usage (important to consider the overall footprint of BC technologies) nor functional metrics (e.g., number of connected peers), and (iii) it only targets private BC platforms.
\item \textbf{Hyperledger Caliper:} A performance evaluation framework called {Hyperledger Caliper} (HLC\footnote{https://github.com/hyperledger/caliper}) which mainly focuses on benchmarking Hyperledger BCs, such as Fabric, Sawtooth, Besu, Burrow, and Iroha. Caliper framework consists of two main components, namely Caliper core (defines system flow), and Caliper adaptor (provides support for integration of various BCs). A predefined configuration file consists of benchmark workloads, and information required for interfacing the adaptor to the SUT is needed before running a test. During the performance testing, a resource monitoring module gathers resource (e.g., CPU, RAM, network, and I/O) utilization data. After the test, a test report containing the values for various performance metrics is generated. Caliper can also monitor server related metrics through Prometheus~\footnote{https://prometheus.io/}. To gain better understanding of the Fabric BC platform, authors in~\cite{8526892} provided a detailed empirical study by using Caliper tool. The study characterizes the performance of BC and highlights potential performance bottlenecks by using a two-phased approach. First phase aims to understand the impact on performance of Fabric (considering TPS and latency metrics) when the configuration parameters, like block size, endorsement policies, number of channels, resource allocation, and state database choice, are used in the test environment. The evaluation results provide various guidelines on configuring these parameters. Moreover, the three performance bottlenecks or hotspots that are identified include (i) verification of endorsement policy, (ii) state validation and commit (with CouchDB), and (iii) sequential validation of associated policies of transaction in a block. Second phase focuses on optimizing the Fabric by considering the obtained observations from performance analysis. Finally, a few limitations of Caliper tool include the lack of network emulation, which is essential for studying the impact of network failure or latency on a BC platform, and the lack of any functionality for resource reservation on scientific testbeds such as Grid’5000. 

\item \textbf{DAGbench:} It~\cite{8814533} is a framework dedicated to benchmarking Directed Acyclic Graph (DAG) DLTs like IOTA, Nano, and Byteball. The performance metrics that it can evaluate include TPS, latency, success indicator, scalability, resource utilisation, and transaction fee. From a system designer's perspective, DAGbench shares an approach similar to BlockBench and HLC tools, i.e., they all adopt a modular adaptor-based architecture. In such designs, if needed, users select or develop the required adaptors to integrate different types of workloads and BC platforms that they want to evaluate. 
\item \textbf{BCTMark:} One of the most recent tools namely \textit{BCTMark}~\cite{02923038}, is a generic framework for benchmarking BC technologies on an emulated network in a reproducible way. To illustrate the portability of experiments using BCTMark, the authors have conducted experiments on two different testbeds: a cluster of Dell PowerEdge R630 servers (Grid’5000) and one of Raspberry Pi 3+. Experiments have also been conducted on three different BC-based systems (i.e., Ethereum Clique/Ethash, and HLF) to measure their CPU consumption and energy footprint for different numbers of clients. The framework provides an abstraction of the underlying physical infrastructure and can be used to deploy quickly on any platform that supports the SSH protocol. In particular, BCTMark provides the following key advantage over other benchmarking tools: (i) playbooks written in Ansible and deployed with BCTMark can be used to deploy an arbitrary number of peers on any testbed that supports SSH connections, while providing the same abstraction over a network, enabling scientists to easily express network constraints and topology, (ii) collects both system metrics like CPU, memory or disk usage (important to consider the overall footprint of BC technologies) and functional metrics (e.g., number of connected peers), and (iii) it can be used with public as well as private BC-based systems.

\item \textbf{BBB:} Recently, authors in~\cite{PAN2020} proposed a benchmarking framework for BCs, namely Boston Blockchain Benchmark (BBB). The BBB design choice allows emulating the underlying networking infrastructure. The participants can then communicate with BC through the emulated network. This provides several benefits to the BBB tool, such as support for fine-grained control over network metrics (e.g., network topology, link latency, and bandwidth) and the integration of various network-related attacks (eclipse~\cite{12831152}, and DDoS attacks). Rather than designing and implementing a network emulator tool, the authors choose to seamlessly integrate the BBB tool with a widely popular (in academia and the industry) emulator called Mininet\footnote{http://mininet.org/.}, which is a battle-tested tool to emulate real-world network scenarios. Although BBB usage is only shown to evaluate Ethereum-based systems, its design is generic, and it is possible to extend BBB to evaluate other BC-based systems. BBB developers claim that the tool is easy to install and use, as it allows every parameter configuration through a simple change of a YAML configuration file. Furthermore, BBB is extensible and configurable.
\end{itemize}
\par Performance benchmarking solutions usually require a standardized test scenario and well-documented workloads as input. However, in case of public BC-based platforms it is not feasible to have an adequate control over a workload and the users participating in consensus process. This makes the use of benchmarking approaches for public BCs challenging. Therefore, there are two potential solutions that could be used for evaluating the performance of public BCs. The first way is to build a test network in a private setting, in a way that it closely represents the public BC's test network, and then leverage the state-of-the-art benchmark tools to evaluate BC performance, by providing artificially designed workloads as inputs. This solution might need the development of a new adapter in a benchmark tool for integrating either the workload or the BC network. In such setting, one must consider the problem of BC scalability, which might arise when the tested private version is implemented publicly. The second way includes monitoring and evaluating the performance of live public BC with realistic workloads as input~\cite{8685888}. For instance, authors in~\cite{8449244} provide a real-time performance monitoring architecture which is based on logging approach. The solution is detailed, and it results in lower overhead and higher scalability, when compared with solutions that use remote procedure call (RPC).

\begin{table*}
\caption {BC-Apps performance evaluation using Benchmarking tools}
\centering
\scalebox{1.2}{
\begin{tabular}{|P{2.1cm}|P{2.1cm}|P{3cm}|P{2.4cm}|P{3cm}|}\hline
\textbf{References} & \textbf{Supported BC platform(s)} & \textbf{Supported performance metrics} & \textbf{Workload details} & \textbf{Limitations}\\ \hline

BlockBench~\cite{1064033} &
Ethereum, Parity, and Hyperledger Fabric &
throughput, latency, scalability and fault-tolerance &
YCSB, Smallbank, EtherId, IOHeavy, CPUHeavy, DoNothing &
insecurity (installed using root user privilege), support only private BC platforms, no testing for faulty behavior \\\hline

Hyperledger Caliper~\cite{caliper} &
Besu, Fabric, Ethereum, and FISCO BCOS networks  &
resource consumption, transaction/read throughput and latency, success rate  &
all types of custom build loads are supported via config files  &
predefined workloads, support only private BC platforms, standard performance indicators \\ \hline

DAGbench~\cite{8814533} &
IOTA, Nano, and Byteball &
resource consumption, throughput, latency, success rate, transaction data size and fee, scalability &
value/data transfers, transaction queries &
specific for DAG DLTs  \\\hline

BCTMark~\cite{02923038}  &
Ethereum Clique/Ethash, and Hyperledger Fabric  &
CPU consumption and energy footprint for different numbers of clients  &
ad hoc load generation based on Python scripts and on an history  &
testing for faulty behavior is not performed, only few performance metrics are covered \\\hline

BBB~\cite{8952155}~\cite{PAN2020} & 
Ethereum, could be extended to other BC platforms & 
tolerance against faulty behavior, network properties (e.g., latency, and bandwidth) affect on BC performance & 
Not specified & 
Does not include failure injection loads, only tested with private BCs \\\hline

\end{tabular}}
\label{T:5}
\end{table*}

\par Apart from benchmarking and live monitoring of BC-based system, the other commonly used PT approaches are based on experimental analysis, i.e., analysis based on self-designed experiments~\cite{8946222}~\cite{8342866}~\cite{8767342}, and simulators, i.e., software tools to mimic the behavior of real-world systems~\cite{1308956}~\cite{13308951}. Next, we discuss some state-of-the-art performance analysis works that use experimental-based performance analysis. 

\begin{itemize}
\item The performance of a private variant of Ethereum BC is studied in~\cite{8342866}, by evaluating Ethereum's two popular clients, namely Proof of Work (PoW) based Geth, and Proof of Assignment (PoA) based Parity. The evaluation results depict that, on average, the Parity is 89.82\% faster than Geth concerning transaction processing rate under various workloads. Authors in \cite{7930225} proposed a technique to predict the latency of private Ethereum (Geth) platform by using a modeling tool called Palladio Software Architecture Simulator~\cite{5959793}. The evaluation also shows a lower relative error in response time (mostly under 10\%). To measure the scalability of  Ethereum BC, authors in~\cite{8705874} used a quantitative analysis approach, in which synthetic benchmarks over an extensible testing scenario are used to measure the transaction throughput. Based on the test outputs, it was observed that Ethereum platform can hardly achieve the three properties, which include decentralization, scalability, and security, concurrently.

\item With the release of long-term support for HLF v1.4 BC platform, it has been evaluated for performance by various researchers. For example, the impact of various workloads on networking infrastructure of HLF v1.4 BC is investigated in~\cite{8946222}. The performance metrics include transactions per second (i.e., throughput), latency, and scalability (measured by evaluating the number of users serviced by the system in specific time period). During the performance evaluation process, various parameters, like number of transactions, generation rate of new transactions, and transaction types (e.g., read or write) were varied to dynamically  change the network load. Similarly, authors in~\cite{8790849} used empirical approach to perform performance evaluation of Sawtooth BC, which is a popular private BC platform under the umbrella of Hyperledger BCs. The metrics considered for performance analysis include consistency (to check if the system produces the same results each time with the same workloads and cloud virtual machine configuration), stability (to check if the system's performance remains stable in the same workloads but under the varying cloud VM configurations), and scalability (to check how the system performance stays scalable with varying parameters related to workloads and cloud VM configurations). It is also observed that by adjusting the configuration parameters, namely scheduler and maximum batches per block, the performance of Sawtooth can be optimised. Finally, based on the results obtained from the empirical performance analysis in~\cite{8946222}~\cite{8790849}, it is concluded that Hyperledger BCs require improvements on geographical scalability, which is usually limited due to its tradeoff with network latency~\cite{Nguyen2019ImpactON}, and size scalability (e.g., \cite{1064033} shows that the system fails to scale with more than 16 peers). Moreover, a major performance bottleneck in scalability is PBFT, the consensus algorithm used in many permissioned BCs. It is because PBFT adopts a communication-bound mechanism for consensus instead of a computation-intensive proof-of-work (PoW) consensus.

\item In the initial implementations of BCs (e.g., Bitcoin and Ethereum), to add transactions in a distributed ledger, multiple user transactions are grouped together to form blocks, which are added in an immutable linked-list type of data structure in the global chain. This process of updating the ledger does not allow the generation of concurrent blocks, thus provides low transaction throughput by limiting the number of transactions that can be added in the ledger per second. Alternatively, in the distributed ledgers that are based on DAG concept, the multiple transactions/blocks can be added simultaneously on different vertices of the directed graph, thus supporting parallel generation and inclusion of transactions/blocks. Inspired from this idea and the need to improve the transactions per second, several distributed ledgers use consensus algorithms that support such concept of transaction generation and addition. For example, IOTA foundation BC uses a cumulative weight approach to confirm transactions, and Markov chain Monte Carlo (MCMC) sampling method to select \textit{tip} (i.e., the vertex in DAG where the newly confirmed transaction can be added) randomly. Other examples include Byteball\footnote{https://byteball.org/Byteball.pdf} system, in which the consensus process relies on a mechanism that selects 12 reputable Witnesses who need to reach to consensus, and Nano\footnote{https://nano.org/en/whitepaper} which adopts a balance-weighted vote technique to achieve consensus on transaction commit. As per the design considerations and theoretical functioning details, the DAG-based DLTs have higher transaction per second. However, it is also essential to identify their performance  bottlenecks or tradeoff between different properties (i.e., scalability and latency). To this end, IOTA scalability in IoT application scenario, consisting of a private network with 40 nodes, has been demonstrated in~\cite{8767342}. The evaluation results show that against the arrival rate of transactions, the processing rate (i.e., TPS) has adequate linear scalability. Furthermore, authors in~\cite{8814533} use their proposed benchmarking tool called DAGbench, to provide a comparative performance evaluation of the above mentioned three DLTs (i.e., IOTA, Nano, and Byteball).
\end{itemize}

\par Next we provide a discussion on two popular simulators that are being used for measuring the performance of BC-based systems.
\begin{itemize} 
\item BlockSim: Three simulators, all named BlockSim (or BlockSIM), were proposed in 2019 for simulating BC-based systems. First, authors in~\cite{1308956} proposed an implementation of a framework called BlockSim, which aimed to create a discrete-event dynamic system model for BC systems that use PoW-based consensus algorithm. BlockSim framework was developed by using a layered approach, and it consists of three layers namely, incentive, connector, and system. With the use of BlockSim simulator, the authors evaluated the performance regarding block creation time metric for DLTs with a PoW-based consensus algorithm, and it provided important insights concerning the block generation process in PoW. To further show the correctness and feasibility of the proposed solution, in their extension study, the authors used verified predefined test cases, where the results of the simulation were compared against the outcome of real-life BCs, like Bitcoin and Ethereum. However, whether the simulator can be extensible or not, this fact is not yet established and needs further research. With the aim to better understand, evaluate, and plan the system architecture and its performance, authors in~\cite{8751320} provide an implementation of their proposed comprehensive simulation tool called BlockSIM. It is open-source and can be used to simulate private BC-based systems. BlockSIM is architect to measure system stability, and transaction per second for private BCs, by running them in various target scenarios. Based on the evaluation results, one can then decide about the optimal system parameters that suit best. This will help to architect a BC-system which achieves the predefined requirements of the application in which it is being deployed. The effectiveness of the BlockSIM is proved by comparing it with private Ethereum network running PoA-based consensus. \newline
Recently, another simulator to evaluate BC projects, with name BlockSim, is proposed in~\cite{8946201}, as a flexible discrete-event simulator. When running BlockSim for Bitcoin and Ethereum BCs, interesting observations related to the performance of BCs were drawn. For example, it was observed that the impact of doubling the size of blocks on block propagation delay is small (i.e., 10ms), but if the communication is encrypted then the delay is greatly affected (i..e, more than 25\%). Later, an extension of BlockSim, namely SIMBA (SIMulator for Blockchain Applications), is proposed in~\cite{9159169}. SIMBA includes Merkle tree as an additional feature on BC nodes and shows that it improves simulation efficiency and allows more realistic experiments that were not feasible before. In particular, the use of Merkle trees provided huge improvements (up to 30 times) in verification time of transaction in a block  reduction, without impacting the block propagation delay. Since the verification of block transactions plays a critical role in the overall computational load of network nodes, improvements in it substantially affect overall performance of network nodes, and consequently, of the whole network.
\item DAGsim: To simulate DAG-based DLTs, authors in~\cite{13308951} proposed a continuous-time, multi-agent simulator, namely DAGsim. The tool was developed by IOTA foundation to mainly test the performance of their BC system (i.e., IOTA) concerning its transaction attachment probability. The analysis results reveled that agents (play a similar role as miners) with lower latency and higher connection degrees exhibit higher chances that their transactions will be accepted by the network. Another multi-agent tangle simulator~\cite{8418134} is developed in collaboration with NetLogo. It simulates random uniform as well as Markov Chain Monte Carlo (MCMC) tip selection, while providing visuals (i.e., GUI) and an interactive experience during simulation. There also exist research works that leverage simulations together with analytical approaches to perform validation and exploration. For example, authors in~\cite{8756973} present a generic DAG-based cryptocurrency simulator. The simulator is built using Python and it was used to validate a proposed analytical performance model. The results reveled that by issuing a transaction with a smaller average number of parents in DAG, it is possible to increase transaction per second. 
\end{itemize}

\par Next, we summarize and provide a comparative discussion along with pros and cons of the above-mentioned different types of empirical performance evaluation solutions. Our comparison considers the generic properties of individual solutions, as well as their assessment suitability while evaluating different BCs platforms. The performance analysis of BC platforms via live monitoring approaches requires a testing deployment network that has high fidelity with real-time production systems, and it also needs realistic workloads and faultloads as its inputs. Although the live monitoring approach is suitable only to evaluate public BCs, it can also be used for benchmarking the private BCs. A common issue associated with evaluating a public BC is that it is difficult to change any parameters to perform several tests. In general, a Benchmarking approach needs a controlled evaluation environment that mainly consists of a SUT network and artificial workloads. The eligible workloads and performance test metrics cannot be easily modified or tuned after selection of a benchmark tool, i.e., these are closely coupled. For example, in Blockbench, network layer parameter's (e.g., network delay) tuning is not supported till date, and it provides support for the evaluation of four specific BC platforms (i.e., Ethereum, HLF, Parity, and Quorum). However, with the design of well-designed APIs, the users can develop additional adaptors to extend Blockbench's functionality to support the evaluation of more private BC platforms. Therefore, compared to the benchmarking approaches, the extensibility of the monitoring approach is lower. Also, the availability of several well documented and open-source benchmarking tools makes them easier to deploy and use than the monitoring approaches.
\par Experimental analysis is one of the most common approaches adopted by researchers to evaluate the performance of their proposed solutions, by using self-designed experiments. It has several similarities with the benchmarking approach, but there are two main differences. First, since the self-design experiments are designed by specifically considering the test requirements and considerations associated with a given BC-App, they provides higher flexibility in evaluating impact factors, and more capability for parameterization. For example, as mentioned before, Blockbench does not support the measurement of an impact of a network delay in HLF, but the same can be evaluated with the help of self-designed experiments. Second, unlike benchmarking tools which are somewhat standardised and their functionality could be extended for evaluating various BC-based solutions, in a self-defined experiment approach the tests are usually dedicated to a specific BC-based solution, thus limiting its extensibility. For both the above performance testing approaches, the difficulty level of the testing environment relies partly on the complexity of SUT and what performance parameters need to be evaluated. In contrast, a simulation-based performance testing approach exhibits high difficulty only in the simulator design and development phase. But, afterwards, the simulator usually renders many advantages, as compared to other approaches. For example, a simulation-based approach is highly extensible, and it can be used to test multiple configurations with different parameter values quickly and cost-effectively. Another inherent benefit of the simulation approach is that it does not depend on the availability of a physical testbed, nor does it need a BC platform. However, the evaluation results obtained from simulation could be quite different from the evaluation results obtained in other performance testing approaches. This, thus, questions the correctness and accuracy of the proposed BC solution that is being evaluated using a simulator. Additionally, there are a few metrics (e.g., transactions per CPU/per memory second/per disk IO, and transactions per network data) for a BC-based system that are hard to evaluate through simulators.

\par Other than empirical approaches, few research works use an analytical modeling approach to perform BC-based systems' performance analysis. Analytical modeling uses mathematical tools to formalize the BC system abstractly, and to solve the resulting models with precision. The model output (e.g., latency expressed as a function of network indicators) gives analytical proof for a BC performance evaluation. Three popular analytical modeling solutions that are being explored for performance analysis of DLTs are queueing models~\cite{3308952}~\cite{10100723}~\cite{8496756}, Markov chains~\cite{8666147}~\cite{8946103}, and stochastic Petri nets~\cite{YUAN2020117}.

\begin{figure}
\centering
  \includegraphics[scale = .26]{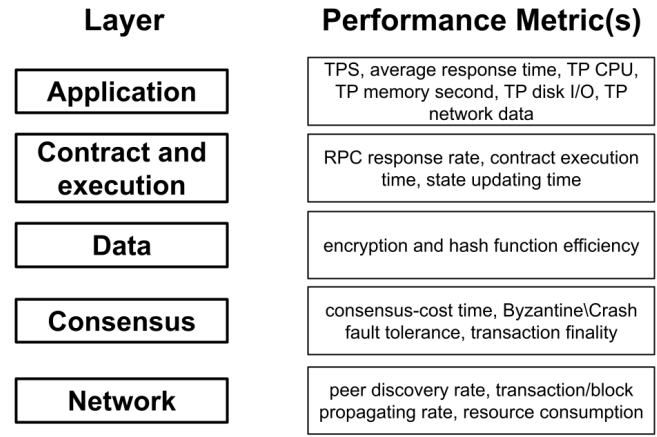}
\caption{Performance metrics at different layers of BC stack}
  \label{pmetrics}
\end{figure}

\par Finally, the metrics evaluated during the performance testing play a huge role in measuring the effectiveness and correctness of the proposed BC-based system. These metrics can be broadly classified in two categories, namely macro (i.e., measured across the whole BC stack), and micro (i.e., evaluated from specific components at different layers of BC stack)~\cite{8449244}. In particular, the performance of the system from the user's viewpoint at the application layer can be measured using the macro metrics. These metrics include transaction per second, transaction processing latency, fault tolerance, scalability, transactions per CPU/memory second/disk IO/network data. From these, the first two are measured frequently over all BCs platforms. On the other hand, the micro metrics include peer discovery rate, remote procedure call (RPC) response rate, transaction propagating rate, SC execution time, BC state update time, consensus cost time, encryption and hash function's efficiency. A well designed workload should be used to evaluate both types of metrics. Moreover, in benchmarking or monitoring approaches of BC performance testing, there are specific workloads that are designed to evaluate the performance of different BC layers. 

\subsection{Security testing}
\label{Security-test}

In this section, we discuss the security and privacy (S\&P) related aspects of BC-Apps, by considering components at each layer of a BC stack (please refer to Figure~\ref{stack}). These aspects should be considered during BC security testing. BC technology supports the concept of security with the help of public-key cryptography and primitives like hash function and digital signature, bit it might give a false impression of security it actually provides. It is because all cryptographic protocols have their limits, and because holistic security includes technology, people, and processes, which are often overlooked in a BC security analysis. The main objective of security testing is to ensure that the BC-Apps are secured against different types of threats caused by viruses and malicious programs injected by malicious entities. Specifically, in BC-Apps, the security analysis and imposed countermeasures should be extremely thorough and responsive due to the unique BC characteristics. For instance, an ongoing transaction cannot be stopped, and thus, the testing process should be effective enough to uncover all potential threats before transaction deployment on BC. Table~\ref{T:6} shows a list of major threats along with their possible countermeasures spread over different layers of the BC stack. During security testing, different test-cases that could lead to these (and other unknown) threats should be prepared to test the SUT. Security testing should also ensure that proper measures are in place to handle such threats if these arise after the BC-App deployment. There exists extensive literature that discusses an array of S\&P threats in BC technology~\cite{9239372}~\cite{8369416}, which should be considered during the security testing and analysis of BC-Apps. 

\par Since most of the threats in a BC-App are associated with the data and processes (i.e., business logic), it is essential to understand the criticality of data and processes. In particular, understanding the sensitivity of the data that is being stored and processed in a BC is needed before one starts the security analysis of such systems. To determine the importance of confidentiality, integrity, and availability (CIA) properties of the data stored in the BC-based system, one should first understand the associated regulatory implications and perform a business impact analysis. During security analysis, a comprehensive threat model that closely reflects the real-world adversary model needs to be adopted. The developers should ensure that the well known threats associated with public key infrastructure (PKI) and application development (e.g., user key leakage, and vulnerabilities or bus in source code) are factored into the security analysis. Additionally, the security threats specific to a BC implementation should be identified. These threats include attacks like hijacking consensus procedure, DDoS, exploiting private BCs and SCs, and wallet hacking. Based on the identified threats, different scenarios representing one or more risks can be listed and evaluated for their likelihood and impact. Finally, based on the identified risks and their impact level, adequate security controls via security analysis procedure should be selected. Moreover, several well-defined security practices like source code review, secure key management, data protection via encryption methods, restricted data access via access control methods, and regular security monitoring, can be deployed. Finally, security improvement techniques specific to BC technology like robust wallet management, authorised ledger management, and secure SC development, should be employed. In particular, it is vital to understand that technology, processes, and people, all are equally important to secure BC-Apps. For example, the DAO hack's damage could have been minimised, if adequate governance structure and incident response process was in place.

\begin{table*}
\caption {Various security threats at different layers of BC-Apps}
\centering
\scalebox{1.3}{
\begin{tabular}{|P{4cm}|P{4cm}|P{4cm}|}\hline
\textbf{Layers and their associated S\&P threats} & \textbf{Potential adversaries} & \textbf{Possible defence mechanisms} \\ \hline
\textit{Application} - false data feeds, CIA and Front-Running attacks, attacks on availability and privacy, malicious trusted execution environment (TEE) or token issuer, censorship, permanent HW fault of TEE &
internal or external attackers (e.g., users, third-party service providers, malware), application/service developers, TEE manufacturers, token issuers, regulatory authorities &
multi-factor authentication, decentralised authority, reputation-based methods, application-level privacy-preserving constructs, HW wallets, redundancy  \\ \hline
\textit{Contact and Execution} - exploiting SC specific bugs &
SC developers, users, external attackers with lightweight node &
safe languages, static/dynamic analysis, formal verification, audits, best practices, mixers, NIZKs, trusted HW, ring/blinding signatures, homomorphic encryption \\ \hline
\textit{Data} - quantum attacks, transaction data tampering attacks &
consensus nodes  &
quantum-resistant cryptosystems, economic incentives, strong consistency, decentralization  \\ \hline
\textit{Consensus} - protocol deviations, violation of  assumptions &
consensus nodes & 
economic incentives, strong consistency, decentralization, fast finality  \\ \hline
\textit{Network} - MITM attacks, availability attacks, network partitioning, routing attacks, DDoS, deanonymization &
Providers of network services &
Redundancy, protection of naming, availability, routing, anonymity, and data  \\ \hline

\end{tabular}}
\label{T:6}
\end{table*}


\par In its default implementation, BC design does not provide support for data privacy, because all the transactions in the ledger are visible to all the network participants. However, pseudonymity is supported by allowing users to use public keys to transact instead of any identifying value. Depending upon the usage application of the BC, a certain level of confidentiality and data protection could be essential considerations during the design and implementation of BC-Apps. One popular way to support data confidentiality is to utilize off-chain storage solutions, in which the organisations can store all or part of their data in local storage. To ensure data integrity for the data stored at local storage, the hash of the data is stored in the distributed ledger. Another approach could involve a fine-grained access control technique to regulate an access to the data stored on the ledger. Most of the solutions will require the use of cryptographic techniques that are still a topic of active research. For example, a version of Zero-Knowledge Proof called zk-SNARKs, allows verifiers to validate a statement about encrypted data, but does not reveal the corresponding decrypted data. Another alternative is to stop broadcasting the transactions in the whole network, and instead, limit the dissemination and visibility to predefined parties in the network. An example of such a design is adopted in R3 Corda\footnote{https://www.r3.com/corda-platform/} BC, which uses an Unspent Transaction Output (UTXO) set model. Finally, the security of transactions is closely coupled with the underlying consensus algorithm  which results an update in the distributed ledger. As advancements in DLT progress, developers have more options to choose which BC platform and consensus algorithm to use for a target application. The selected design elements, along with their rules, will set various networking parameters such as transaction speed, latency, and scalability. Therefore, during the requirement analysis, or before developing a proof-of-concept, it is vital that developers and security engineers carefully evaluate the algorithms and protocols, to identify the ones that are best suited for their specific BC-App.

\par Consensus protocols are critical to BC security, therefore, understanding potential threats to their security is essential to securing the BC. To this end, a large number of consensus protocols have been proposed~\cite{180905613}~\cite{13355458}, some of them with the aim of improved security against many threats and high performance in large scale networks. However, the adversaries are continuously targeting the consensus protocols, either to destabilise it or to gain financial profits from the target BC-App in which it is being used. Therefore, the security testing should cover all the possible test cases and adversary models, to check, if the consensus protocol used in the target BC implementation system could be exploited for vulnerabilities leading to other security threats. In particular, the consensus protocol should be able to tolerate or quickly recover from faults (caused by adversary or system generated), to ensure that it finds consensus and complete transactions even in a sub-optimal network topology. Also, the designers of BC-Apps should take precautionary measures (e.g., economic incentives, strong consistency, decentralization, and fast finality) the reduce the risk to consensus protocols. While performing security testing for consensus protocols against well known threats, it should be ensured that the protocol can satisfy the following three essential properties: (i) consistency (i.e., the network peers should agree on a proposed value to reach consensus in certain time limit), (ii) transaction censorship resistance (i.e., resilience to malicious nodes blocking genuine transaction), and (iii) distributed denial of service (DDoS) resistance (i.e., resilience to malicious nodes launching DDoS attacks on consensus algorithm).

\par As discussed by authors in~\cite{ferdous2020blockchain}, a consensus algorithm must satisfy a number of security properties, and the same should be tested during its security analysis. These properties are as follow: (i) Authentication, it implies whether nodes participating in a consensus protocol need to be properly verified/authenticated, (ii) Non-repudiation, it signifies whether a consensus protocol satisfies non-repudiation, (iii) Censorship resistance, it implies whether the corresponding algorithm can withstand against any censorship resistance, and (iv) Attack vectors, it implies the attack vectors applicable to a consensus mechanism. The attack vector can be further divided into the three threats against which a consensus protocol should be tested. These threats include: (i) Adversary tolerance, which signifies the maximum byzantine nodes supported/tolerated by the respective protocol, (ii) Sybil protection, in which an attacker can duplicate his identity as required in order to achieve illicit advantages. Within a blockchain system, a sybil attack implicates the scenario when an adversary can create/control as many nodes as required within the underlying P2P network to exert influence on the distributed consensus algorithm and to taint its outcome in her favour, and (iii) DoS resistance, which implies if the consensus protocol has any built-in mechanism against DoS attacks. Finally, securing the consensus protocol should not come at the cost of low performance, i.e., it should not adversely impact the latency, throughput, and scalability, and a suitable trade-off should be considered. This trade off strongly depends on the target application requirements in which the BC is integrated, and the same should be taken into consideration during the security analysis.  

\par Depending upon its implementation nature (i.e., private or public), a BC networking infrastructure is based on either private or public networks. A private network uses centralized administration and it supports features such as low latency, user and transaction privacy, and compliance with regulatory obligations (e.g., HIPAA for healthcare data). Private networks inherently provide authentication and access control, and have full control over communication routes and network resources used, enabling suitable network topology regulation about the given requirements. The network administrators can apply fine-grained access control techniques to implement the security principle of minimal exposure. This way, the insider threats in a local network can be minimized. Authors in~\cite{9239372} observed that internal and external attacks are the specific security threats to permissioned BCs. For example, permissioned BCs use centralized access control that can be attacked by an external attacker by exploiting a network or system vulnerability. Such an attacker is even easier to launch for the internal attackers, as they might already have the required privileges or can get them by exploiting certain vulnerabilities in systems, network, or organizations involved in the BC-App. One result of such an exploitation could be that the internal attacker can add malicious miners (i.e., nodes running consensus algorithms) or remove legitimate ones from the network. Such change in the network  will result in increasing the adversarial hash rate (aka consensus power) demonstrated at the consensus layer. Moreover, the attacker could launch many attacks, such as double spending, and attacks on violation of protocol assumptions, that can be benefited with the increased hash rate at consensus layer. Therefore, it is vital that security testing considers all these possible attacks before deploying the permissioned BC-Apps. Testing the security of permissioned BC is easy compared to its counterpart, due to the controlled environment's availability. On the other hand, the public BCs are and should be more secure by the implementation, as they are built to deal with many unknown entities (including the malicious ones) participating in different activities of the BC ecosystem.  

\par BC-App components such as peer nodes and APIs use underlying private or public network for communication. Depending upon the type of BC-based solution implemented, the peer nodes and their associated roles can differ in different implementation scenarios. These nodes use participants’ own (in case of public BC) or an organisation's (in case of private BC) networking infrastructure to communicate with the BC network. These networking infrastructures should be equipped with required fundamental security controls and measures, e.g., penetration testing, periodic vulnerability checks, log monitoring, endpoint vulnerability testing, and security patch or update management. The lack of defense mechanisms could lead to the compromising of client nodes, and a single compromised node can result in loss of assets of the client associated with that node in case of public BCs. While in private BC scenario, such malicious node will remain undetected, and violating the privacy of the attacked client node by eavesdropping on the transactions performed by the node. The considerations needed to protect the client nodes from such attacks are align to the BC security recommendations set out by Gartner\footnote{Gartner, Evaluating the Security Risks to BC Ecosystems, March 2018.}. The recommendations include steps like taking a holistic view of security, and ensuring the risks are evident at the business, technical, and cryptographic levels. Moreover, same as with any technology implementation, all BC-based projects need to be analysed for their readiness to handle a security threat, and with incident response plans in place to handle critical security events during the BC life-cycle. On the one hand, there already exists a large number of security threats at different layers in the BC stack~\cite{9239372}, and its integration with an application will further increase this attack surface. But, apart from the security analysis of SCs, there are none or little support, in terms of tools or techniques available in the state-of-the-art for performing security testing of various components of the BC-Apps (e.g., consensus protocols, peer nodes, and integrating endpoints). A point to be noted is that based on the research articles in literature, there exists a large number of security threats to standalone BC systems as well as BC-Apps, but the security incident types occurred in practice are significantly lower, mainly at consensus and application layers. At application layer, a large percentage of security incidents are caused by exploiting a centralized component by external or internal adversaries, while at consensus layer, majority of incidents are caused via temporary violations of protocol assumptions by 51\% attacks.  

\subsection{API and Interface testing}
\label{api-test}

In a BC-App, the users of the application interacting with the BC will have APIs in use to connect to the BC. A BC-App ecosystem comprises of different components and all these components must be connected. Also, it is crucial that the different APIs associated with these components are tested for their compatibility with each other. API testing plays a major role to ensure that the backend is functional. Like any other API, the APIs in BC-Apps also need to be tested for any potential flaws like unauthorised access, encrypted data in transit, and cross site request forgery. For example, APIs developed to interact with BC need to be checked for errors like starting an automatic call over a large number of transactions. Such a common mistake in the API development could be very costly, especially in BC networks where transaction processing has a cost (e.g., gas in Ethereum BC). The API tests need to check that all the interactions between applications/users and the backend (BC network) in the BC-App are as per the predefined specifications, and the performance of the interaction is correct and smooth, i.e., application can process and format API requests optimally, and they can check and verify that all API requests/replies from the backend are handled correctly. Finally, testing APIs helps in BC's block verification process. It is because each block information has a unique hash that changes when there is any change performed in the block, and the APIs that fetch and verify these hashes can be validated through API testing.

\par All the APIs would require thorough testing for functionality to ensure that there are no functional issues and that the service integration works seamlessly. In a practical BC-App scenario, there exist two types of interfaces. First is between the BC and the application in which API is being integrated, and second is between the various components of a BC-based system. The former is more challenging to test in an efficient manner, as it requires the significant knowledge of both domains and specialised testing tools that can work with both these domains. For instance, usually the users of DApp interact with the BC using a web application browser. In the existing test scenarios, the methods that test the web applications~\cite{6032496} only consider browser-side code, while the SC analysis tools~\cite{Kalra2018ZEUSAS} consider only the SC code. This approach of independent testing makes it challenging to use these techniques as they are in the DApp setting. Therefore, in practice, the testing procedure should consider the fact that the testers need to work together to understand the interfacing between the browser program (e.g., JavaScript code) and BC programs (i.e., SCs). Finally, API testing should ensure that APIs are secure, simple, and that they provide high performance for adequate usability of the underlying system.

\section{Discussion and Future Research Directions}

In this section, we provide a brief discussion that includes the advantages and limitations of the current state of testing along with the lessons learned based on our comprehensive study on BC-Apps testing. In particular, this discussion is derived from our detailed survey on state-of-the-art testing efforts (i.e., testing tools and techniques) that we have presented in Section IV. Finally, we also mention a few future research directions that need attention from the research community which is exploring the techniques and tools for effective and efficient testing of BC-Apps. In particular, the future research directions are derived from the existing challenges (refer Section III) and efforts (refer Section IV) related to testing BC-Apps. 

\par \textbf{SC Testing:} Despite the increasing interest shown from its usage in a large array of applications, SC development remains somewhat a puzzle to numerous developers, primarily due to its unique design. Therefore, it is required that the research investigates the critical questions related to the development process of SCs. These questions could be as follows: (i) are there any differences between SC-based software and traditional software development procedures, and (ii) What type of challenges the SC developers face. We broadly classify the SCs vulnerability testing tools in three categories in Section III-A, namely static, dynamic, and formal verification. Authors in~\cite{8782988}~\cite{3395363339} provide a comparison of these categories concerning their performance, coverage of finding vulnerabilities, and accuracy, which could be referenced while selecting a testing approach for SC testing. Automation tools implementing static and dynamic analysis methods are convenient to use to analyze vulnerable SCs. However, they detect only their specifically defined vulnerable patterns, limiting the testing expanse, as defined patterns are occasionally exhaustive. In contrast, formal verification approaches use theorem proving methods to validate SCs' fitness properties using their interpreted proofs.

\par As mentioned in Section III-A, researchers have proposed many tools and techniques to statically discover bugs and vulnerabilities in SCs. Nevertheless, there are many recent works and incidents that reported various vulnerabilities in SCs. This questions the efficacy of the state-of-the-art tools and techniques used to detect the vulnerabilities in SCs. It could be because most of the static and dynamic analysis tools have been evaluated either on developer generated data-sets and inputs, or on data-sets of contracts with a limited number of bugs. Mainly, the existing solutions do not provide the required level of code coverage that could uncover all the threats, while false positives and false negatives remain high. Moreover, empirical studies of software defects have shown that it is possible to detect many defects by using static analysis tools, but this is true only in theory due to limitations of the tools~\cite{6494905}. Recently, authors in~\cite{3395363339} focus specifically on the undetected bugs (i.e., false negatives), but also consider the false-positives. 

\par Formal testing is considered as an important component of software development life cycle (SDLC), which tests the software behaviour and performance against its predefined specifications and requirements, by testing it for a predefined set of possible input conditions. However, in the SC development process, formal testing has been significantly overlooked. Although it is quite possible to perform a transition of rules and guidelines of formal testing from a standardised software (i.e. well-studied and documented) to SC-based software for its verification. For example, using a testing environment in truffle, one can create formal test cases for Solidity SCs based on certain mathematical logic and rules, and these can be executed to check SC properties. Therefore, different approaches using formal methods should be investigated for their potential to effectively test SCs. For instance, property-based testing~\cite{7899079}, which aims at verifying program's properties that should remain true on all the inputs generated from a wide range of possible inputs for that program. Similarly, another promising research direction for detecting bugs and vulnerabilities in SC is automated formal verification, which checks the functional correctness of SCs. Although formal methods are quite effective for the verification of SCs, a SC could still contain bugs, while the formal verification framework could incur large bug detection time, be expensive and highly complex. In particular, one can never be certain that specified properties used during formal testing will detect all undesirable outcomes of a SC. If some important properties are forgotten during the verification, the SC could remain buggy.

\par Recently, authors in~\cite{8847638} concluded that developers are facing several significant challenges during SC development, based on findings obtained from interviews and a survey. These challenges include (i) there is no practical way to assure the security of SC code, (ii) existing tools for development are still immature and exhibit limited functionalities, (iii) the programming languages and the virtual machines still have several limitations, (iv) performance obstacles are difficult to manage under a resource-constrained running scenario, and (v) online resources (including advanced/updated reports and community support) are still insufficient.

\par  \textbf{Performance Testing:} As an important component of BC research, performance evaluation plays crucial role in improving BC-Apps. Although numerous tools and techniques for BC performance evaluation have been proposed and implemented in the literature, as mentioned in Section III-B, only few of them have been well analysed and evaluated. Due to the unavailability of standardised interfaces while running workloads, comparative analysis between various BC platforms is challenging. Specifically, for BC systems that differ in consensus protocols and data structures that they use. Apart from basic functionalities in the most popular BC platforms (e.g., HLF and Ethereum), there is a requirement for performance testing approaches to evaluate new functionalities that are continuously being proposed by the research community. For example, sharding approaches~\cite{10319889} are being considered as a viable solution to BC scalability, and the same have been implemented in many BCs. However, such shard-based BCs are not compared for their performance, in particular. Therefore, it is not clear how the use of sharding could impact the performance of the target BC system. Moreover, the impact on performance of BCs implementing different solutions like sharding vs DAG, and off-chain vs side-chain also needs to be comparatively evaluated. Additionally, it would be beneficial to evaluate a BC performance by combining empirical and analytical performance evaluation techniques.

\par  \textbf{Security Testing:} The lack of best practices along with innovative tools/techniques for security testing of BC implementations are the key reasons that make BC security a significant challenge. In BC-Apps, vulnerabilities that lead to security threats are not limited just to the BC and SCs. Other aspects are susceptible to vulnerabilities, such as governance, compliance, human errors or misunderstanding, and assert at risk, which should also be evaluated. Moreover, the BC implementations should embed security mechanisms specifically targeted to various components that lie at different layers of the BC stack. This should be done from design phase, instead of putting security patches afterwards when a vulnerability is detected. Such an approach to security can hopefully provide a robust defense, and make a BC platform cyber-resilient. We insist that for each layer of our BC layered model, we consider secure cryptographic primitives with recommended key lengths based on existing standards~\cite{Pritzker2015}~\cite{Cavanagh2015}. Examples involve secure communication (i.e., network layer), the use of digital signatures that are based on private keys for signing the transaction (i.e., consensus layer), and login credentials management for BC-oriented services (i.e., application layer).

To provide end-to-end security in BC-Apps, it is required to ensure that all the individual components of BC (e.g., SCs, consensus algorithms, peer nodes, peer-to-peer network, and an offchain storage system) are tested for vulnerabilities~\cite{9239372}. Moreover, new threats arising from the integration of BC technology with the target application should be analysed as well. Based on the state-of-the-art presented in Section IV-C, we observed that many security attacks happened for the application layer due to exploiting a centralized element by external or internal adversaries. In contrast, in the consensus layer case, many attacks happened due to a temporary breach of protocol hypotheses by 51\% attacks. Therefore, further research is required to propose solutions to address these threats at both the layers. Moreover, to manage safe and correct software at each of the layers, alike to the contract and execution layer case, developers should employ verification tools, code reviews, testing, audits, known design patterns, and best practices. We believe that as BC-Apps implementations mature with time, new security threats will be discovered, but at the same time there will be a more mature security testing strategy and a rich set of BC-specific security testing tools available. However, at present, it is safe to conclude that there is a lack of comprehensive security testing frameworks that could be used to perform adequate testing of BC-Apps. 

\subsection{Future Research Directions}

\begin{itemize}

    \item Guiding procedures for BC-Apps testing - There is a lack of standardized best practices in developing BC-based systems which can alleviate the open challenges of testing such systems. For instance, at present, SCs follow a non-standard software development life cycle, according to which delivered applications can hardly be updated, and bugs can only be resolved by releasing a new version of the software (i.e., hard fork). Therefore, the most important research direction for BC-based testing is to establish a set of guiding procedures that could enable developers to carry out a testing strategy specific to BC-Apps. It will help them to cover the essential steps needed to ensure that the developed BC-App will work as predicated in the real-world environment. The best practices defined in the vast literature of software engineering for testing software systems help, but only partially. It is because BC-based software systems exhibit some unique properties (e.g., immutability, and transparency) and new components (e.g., SCs, and consensus algorithms) that need further investigations before defining standardized testing guidelines. \newline
    With the rapid implementation and deployment of BC-based solutions by various enterprises, the use of standard best practices for reviewing a BC source code are particularly encouraged. Emphasis should be given to performing peer-review and testing software by external independent team prior to its release. One important step that the guidelines should emphasize is the usability testing of the target BC-App implementation. It is to improve the overall quality of the implemented software for its clients. Current research shows that user experience when interacting with BC-based software can be a significant concern~\cite{101007978}. For example, users must pass a lot of hurdles to interact with these software systems, if the decentralization tenet of BC is kept intact. However, such hurdles are hard to demonstrate and articulate, due to the lack of proper measuring tools for such purposes. Guidance and documentation can only get so far. Therefore, an interesting research direction would be designing and engineering the client-side of BC-based software such that it can honour the decentralization without compromising user experience (i.e., usability).       
    
    \item Compliance testing - The inherent features of BC, such as transparency and immutability, could cause S\&P related threats to data owners, when their personal data is being processed and managed via BC-based systems. For instance, if there is a requirement for the BC system to comply with the ``right to be forgotten'' regulations, then this is conflicting with the potential immutability of data feature that the BC provides. Such issues make compliance testing for BC-Apps important, because failing to comply with regulatory bodies (e.g., GDPR or HIPPA) can have serious consequences, such as fines, negative press, revenue decline, and even jail time. The compliance testing will also ensure that all data privacy-related risks stated in the associated regulation acts are taken into consideration during the design of the target BC-based system. Compliance testing checks that there exist no flaws in the BC-based system design that could lead to non-compliance with regulations or confidentiality agreements governing data. For instance, the following could be checked during the compliance testing, related to data privacy regulations: (i) does the application involve personally identifiable information (PII) or confidential freight data, and (ii) do the application requirements allow on-chain data storage, or data must be stored off-chain. The importance of compliance testing and the lack of any compliance testing strategy or a tool for BC-Apps shows that there is a huge research gap in this area that needs to be filled.
    
    \item Ecosystem or third-party risks analysis - Compared to the BC-Apps where BC technology is integrated within the target application, the standalone BC platforms (such as Bitcoin, and Ethereum) have proven secure till now. However, the security of a BC-based solution relies on all the applications that are part of the ecosystem in which the BC is integrated. Often, such an ecosystem consists of multiple organisations and third-party service providers (e.g., SC developers, and wallet and payment platforms). This heterogeneity of autonomous organisations makes testing the ecosystem as a whole challenging. For instance, different organisations might use different type of devices, communication protocols, and security protocols. Specifically, an organization's BC-App consisting of third-party BC solutions and platforms is as secure as its weakest link across all the technology provided. The security considerations of a public BC differ from the security requirements of each organisation or a service provider taking part in the BC-App ecosystem. Therefore, to avoid vulnerabilities caused by third party services, it is required to do a thorough vetting of parties involved in the ecosystem. The vetting phase should be followed by a comprehensive testing that could ensure that the performed security tests cover the risks associated with the usage of third party solutions that are integrated in the BC-App. 
    \item Testing automation - Since the deployment of BC-Apps is still in its early phases, there is a lack of automation tools for designing and testing these implementations. The lack of such tools makes the testing expensive and time-consuming, thus disincentivises the BC-App developers to go through the testing phase. Moreover, the highly competitive market to provide BC-based services further adds fuel by creating a race condition between different BC-based service providers. Therefore, there is a huge demand for the deployment and testing automation tools for BC-Apps to make testing faster and more efficient. For instance, the BC network deployment process is usually complicated and therefore should be automated, so that the time and resource consumption of testing can be reduced significantly. Some deployment automation utilities\footnote{https://www.ansible.com} exist on the market, typically as a part of a blockchain-as-a-service offering. However, they are limited in terms of the BC platform and hardware infrastructure that they support. Moreover, they fail to capture high-level design decisions that stem from the design time, and thus do not represent the high-fidelity testing setups when compared with the production environment. \newline
    Manual test generation is likely to form an important component, but inevitably is limited, therefore, there is a need for effective automated test generation (and execution) tools. To this end, researchers have already started to work towards the creation of automation tools for SC testing~\cite{8946219}~\cite{3404366}. Moreover, there are few automation tools for performance testing~\cite{8814533} as well. However, these tools are at early stages and do not support the automation of all the required functionalities and operations~\cite{3391195}~\cite{9129732}. For example, Caliper tool takes a configuration file as input, to represent the workload, and falutloads to evaluate the performance of the system, but it does not allow the SC functions to interact with the fabric during the evaluation process. Thus, it does not truly automate the performance analysis process when complex SCs are involved that need to interact with the BC system during its execution. Therefore, automation with scripts or specialized IT automation tools, and work towards a model-based approach, needs attention from BC software development community. 
    
    \item Testing endpoint vulnerabilities - BC enters the market as a new technology and it rapidly finds its usage in huge number of application domains. This provides an incentive to developers and BC-based service providers to be first to release their solutions, often at the risk of deploying partially tested code on BC-Apps and sometimes on live BCs. For any new technology, its endpoints are most vulnerable, thus an easy target for the adversaries to launch malicious attacks. The endpoints, such as digital wallet, a specific device, or any user-side application, are interfaces for users/clients to connect with the BC-App solutions. If an adversary can compromise (i.e., get possession of a user's password or private key, or get physical access to the devices) any of these endpoints, it can get access to the user accounts, unless enhanced security measures, such as multi-factor authentication or fine-grained access control, are in place. Such malicious access, if successful, can put everything in the user account at risk. With such access to user accounts, the attacker can misuse the account without rising any external alarms or leaving signs of any abnormal behavior. In private or centralised systems like banking, there are possibilities to detect and even correct (e.g., reverse a transaction) a malicious transaction. However, the decentralised and immutable nature of public BCs does not support such corrections, and the same applies to private BCs up to a certain point. Therefore, thoroughly testing the interfaces through which users access the BC infrastructure is of high importance. To ensure that the client-side applications are secure, progress has been made as protective measures, such as the use of cold wallets, often along with the hardware security models (HSMs), which are being implemented by various companies. Unlike hot wallets that need internet connection and store all the account information (e.g., password and private keys) in an online storage, the cold wallets work in an offline mode, thus making them hard to compromise. In summary, it is essential to carefully check and address all end-point vulnerabilities that exist at the integration points of different components of BC, and between the BC and endpoints of the application in which it is integrated. Performing an adequate testing in BC-Apps is particularly challenging due to a vast knowledge set one needs to have.  
    
   
\end{itemize}

\section{Conclusion}
In this paper, we provide a comprehensive study on testing of BC-Apps, which includes the challenges it faces, the techniques and tools available to test various BC components, and the different types of testing required for it. Moreover, we identify a set of research gaps that need attention from the research community working on the topic. As concluded from our study, the key component requiring extensive and rigorous testing are SCs, and the same have received a lot of attention from researchers that have proposed different tools to test them for bugs and vulnerabilities. Such tools aim to automate BC testing, improve code coverage, and achieve low false positives and negatives. Next, there exist a few tools that provide support for performance testing, but these tools are at their primarily phases, and require further improvement and new  test features. The efforts toward the security testing of BC-Apps are not addressed adequately. There are research works that address the security of individual BC components, but tools that could access the security of the whole BC stack (i.e., end-to-end threat detection) do not exists yet. Finally, techniques to test the performance and security of consensus algorithms and BC nodes need to be explored. The issues related to testing of applications that use SCs and BC technologies raise huge concerns to developers. It is because the rapid usage of these technologies in industries is currently worth billions of dollars. Hence, the testing of these technologies needs testers from multiple research domains (e.g., distributed systems, new coding languages, formal verification and validation methods, and cryptography) to work together to perform inter-domain research activities. Achieving a significant progress on these issues would be hard if the research challenges and gaps mentioned in this paper are not addressed properly. To this end, we hope that the testing challenges and research gaps highlighted in this work will help the research community to coordinate and work together to improve SCs and BC-based solutions in general.

\balance

\end{document}